\pdfoutput=1
\documentclass[a4paper,11pt]{article}
\usepackage{amssymb}
\usepackage{amsmath}
\usepackage{jheppub}
\usepackage{amsfonts}

\affiliation[a]{Facultad de Ciencias, Universidad Arturo Prat,\\
Avda. Arturo Prat 2120, Iquique, Chile}
\affiliation[b]{Instituto de Ciencias Exactas y Naturales,\\
Avda. Playa Brava 3256, Iquique, Chile}
\emailAdd{patsalgado@unap.cl}
\abstract{We propose a first-order gauge framework for de Sitter relativity in which the
pseudo-radius $l(x)$ is promoted to a spacetime field, locally breaking
$SO(4,1)\longrightarrow SO(3,1)$, with $l(x)$ acting as a geometric
compensator field. This promotion modifies the Strong Equivalence Principle
so that the local vacuum tracks the local matter distribution rather than
remaining fixed. The resulting action is $SO(3,1)$-gauge invariant but not covariant under the full translational sector of $SO(4,1)$. We make $l(x)$ dynamical through a topological Nieh-Yan coupling. Once the algebraically determined torsion is eliminated, this coupling adds one propagating scalar degree of freedom to General Relativity. The three resulting field equations, for $l(x)$, the Lorentz connection and the vierbein, are shown algebraically to be equivalent to the tensorial splitting $G_{(T)}^{\mu\nu}-G_{(K)}^{\mu\nu}$. The gauge
Bianchi identities reveal that matter spin density sources spacetime torsion
while the axial component of that torsion alone sources the local variation
of the cosmological parameter $\Lambda(x)=3/l^{2}(x)$. On a Friedmann-Lema\^{\i}tre-Robertson-Walker background with trace and axial torsion, the torsion equation fixes the axial torsion in terms of $\dot{l}$, and the $l$-equation becomes a second-order equation. The de Sitter scale then behaves as a quintessence field with an inverse-square (Ratra-Peebles) potential, $V=\Lambda =3/l^{2}$. The reduced system is
consistent with standard matter conservation, $l(t)$ is bounded away from
zero, and at late times $\Lambda(t)\propto t^{-2/3}$ relaxes to zero while
the expansion accelerates with $w_{\mathrm{eff}}\to-1$, without requiring
any chiral spin density.}
\makeatletter\gdef\@fpheader{\ }\makeatother
\begin{document}

\title{\boldmath A First-Order Gauge Approach to de Sitter General Relativity%
}
\author[a,b]{P.~Salgado}
\maketitle

\flushbottom

\section{Introduction}

\label{sec:intro}

Modern theoretical physics rests on two pillars that are
individually successful but not yet unified. Quantum Field Theory describes the microscopic behavior of matter through gauge symmetries in a flat Minkowski arena, while General Relativity describes gravity as a manifestation of spacetime curvature. In standard Einsteinian gravitation,
the connection between these two frameworks is governed by the Strong
Equivalence Principle (SEP) \cite{1}. The SEP states that, at any point in space and time, a local coordinate system can be chosen in which the effects of gravity vanish locally. A freely falling reference frame is an example. In such a frame, the laws of physics reduce to those of Special Relativity. Formally, this requires the local tangent space to be flat
Minkowski spacetime, structured by the kinematical symmetries of the Poincar%
\'{e} group $ISO(3,1)$. The gauging of this group is the historical origin of the gauge-theoretic approach to gravity \cite{2,3,4}. Its extension to a dynamical, torsionful connection, the Einstein-Cartan-Sciama-Kibble ($U_{4}$) framework, has been developed in Refs.~\cite{5}-\cite{9} and provides the broader context of the present de Sitter generalization.

This paradigm has strong predictive power, but it faces structural and cosmological challenges when confronted with the global
evolution of the Universe. One example is the cosmological constant problem 
\cite{10,11}. When the Einstein field equations are amended
with a cosmological term $\Lambda $ to account for the observed accelerated
expansion of the universe \cite{12}-\cite{20}, the second contracted Bianchi
identity ($\nabla _{\mu }G^{\mu \nu }=0$) imposes the restriction $\partial _{\mu }\Lambda =0$. As a result, $\Lambda $ is constrained to be a global
constant across all space and time.

This mathematical rigidity leads to a problem. Quantum field theory
interprets $\Lambda $ as the energy density of the vacuum, a calculation
that overestimates the observed value by roughly $120$ orders of magnitude.
At the same time, describing the dynamics of the early universe, such as the
primordial inflationary phase, requires a cosmological parameter that varies
over time \cite{21}-\cite{27}. A number of approaches have been proposed to
address this, including the introduction of scalar fields \cite{28,29,30} or
modifications of the geometric action functional \cite{31}-\cite{59}. These
approaches generally retain the flat Poincar\'{e} kinematics of the tangent
space while modifying the dynamical gravitational sector.

This paper explores an alternative approach, based on the framework
developed by Aldrovandi and Pereira \cite{60}-\cite{71}. Rather than
modifying the gravitational action or introducing new fields, we examine the
Strong Equivalence Principle at its kinematic level, following these
references. The central question is whether the local
kinematical background of spacetime should be modeled on the flat Minkowski
space of the Poincar\'{e} group, or on the curved geometry of the de Sitter
group $SO(4,1)$.

The idea of allowing the de Sitter radius itself to vary, $%
l\rightarrow l(x)$, has been considered previously, but the conclusions
reached depend on which gravitational action is used to
implement it. This distinction is directly relevant to situating the present work.

Ref.~\cite{72} promotes $l\rightarrow l(x)$ within \emph{teleparallel} gravity,
i.e.\ a curvature-free, torsion-based formulation of the gravitational
sector. There, the resulting cosmological function $\Lambda
(x)$ is given its own dynamics, governed by a wave-type equation sourced by the trace of the torsion and analogous to that of a nonminimally coupled scalar field. It also contributes kinematically to the geodesic deviation equation for free-falling particles. No equivalence to unmodified General
Relativity is claimed or expected.

Ref.~\cite{73}, by contrast, promotes $l\rightarrow l(x)$ within a \emph{%
curvature-based}, MacDowell-Mansouri-type $SO(4,1)$ gauge action, the same
type of construction adopted in the present work. There, the
resulting action is shown to be invariant under the Weyl rescaling $g_{\mu
\nu }\rightarrow f^{2}g_{\mu \nu }$, $l\rightarrow fl$, and is therefore
equivalent, point by point in this gauge sense, to General Relativity with $%
l $ absorbed into the definition of the metric. This is verified on an FLRW
background, where the entire family of solutions with time-dependent $l(t)$
reduces, after the Weyl map, to the standard Friedmann equations. Ref.~\cite{73} also notes that this equivalence is a statement about a \emph{%
proper} Weyl transformation. At points where the rescaling factor $f=l_{P}/l$
vanishes or diverges the map becomes singular, and inequivalent, physically
distinct solutions (e.g.\ traversable wormholes) can arise. The equivalence
with General Relativity therefore holds as a local gauge identity away from
these degenerate points, rather than as a global statement valid everywhere
on the manifold.

Because the present manuscript adopts the same
MacDowell-Mansouri-type action (namely a four-dimensional Lanczos-Lovelock
action) as \cite{73} (Eq.~(\ref{6}) below) rather than the teleparallel one
of \cite{72}, the equivalence result of \cite{73} applies directly to the
purely geometric part of our construction. Without the Nieh-Yan coupling 
\cite{74} introduced in Sec.~\ref{sec:nieh-yan-eom}, promoting $l\rightarrow
l(x)$ within this action does not introduce new propagating dynamics beyond
General Relativity in the regular sector. The Nieh-Yan coupling changes
this: since the torsion it sources depends on $\partial l$, eliminating the
torsion endows $l(x)$ with a kinetic term, and the theory describes General
Relativity plus one propagating scalar field (Sec.~\ref{sec:6.2}). A more general treatment encompassing both cases is given in Ref.~\cite{75}. There it is shown that a whole polynomial family of $SO(1,4)$
actions for a dynamical compensator field reduces to General Relativity in
the limit where its norm approaches a positive constant. The generic members
of the family exhibit new phenomenology (quintessence-like dynamics,
propagating torsion, or signature change) in the irregular regions where
that limit fails to hold.

The contributions of the present work are:

$\left( i\right) $ an explicit first-order, Cartan-geometric derivation of
the torsion sector associated with a variable $l(x)$, expressed in the language of differential forms rather than in the metric formulation used in Refs.~\cite{60}-\cite{73}. This language is the natural setting for an eventual coupling to fermionic matter.

$\left( ii\right) $ a topological Nieh-Yan coupling that supplies $l(x)$
with an independent equation of motion.

$\left( iii\right) $ the complete reduction of the resulting system on a
Friedmann-Lema\^{\i}tre-Robertson-Walker background with torsion, which
shows that, once the torsion is eliminated, the de Sitter scale behaves as a
quintessence field with an inverse-square potential.

Replacing the Poincar\'{e} group with the de Sitter group as the local
kinematical symmetry changes the Strong Equivalence Principle without
modifying the underlying Einsteinian dynamics. Under this modified SEP, a
freely falling observer experiences a local de Sitter tangent space
characterized by a pseudo-radius $l(x)$, which acts as a kinematic length
scale. Because the de Sitter group is transitive, the cosmological term $%
\Lambda(x)$ is not introduced as an external addition to the theory; it
appears within the structure constants of the local tangent algebra,
representing the kinematical curvature of the background.

Previous treatments of de Sitter relativity have generally relied on the
metric formulation in coordinates (see \cite{60}-\cite{71}). This
formulation presents mathematical difficulties associated with the
nonlinearity of the metric tensor, particularly when separating the
dynamical curvature of gravity from the kinematic curvature of the de Sitter
background. To address these issues, we develop a first-order gauge approach
to de Sitter relativity using the language of differential forms and a
modified four-dimensional Lanczos-Lovelock action \cite{76,77}.

When General Relativity is formulated over a locally curved de Sitter
background instead of a flat Minkowski vacuum, the geometric description of
spacetime is no longer strictly Riemannian and is instead formulated using
the language of Cartan geometry. As established in Sharpe's
generalization of Klein's Erlangen program \cite{78}, Cartan geometry
provides a mathematical framework for describing manifolds that are locally
modeled on homogeneous spaces rather than flat vector spaces. This framework combines the spatiotemporal translation sector and the Lorentz rotations
into a single algebraic connection form.

\section{Gauge fields for spacetime de Sitter symmetries}

The spacetime symmetry group connects inertial coordinate systems. Following 
\cite{79}, we will suppose that infinitesimally separated inertial frames
are connected by the transformation 
\begin{equation}
x^{\mu }\longrightarrow x^{\prime \mu }=x^{\mu }+\sqrt{x^{2}}\varepsilon
^{\mu }+\lambda ^{\mu \nu }x_{\nu },  \label{p1}
\end{equation}%
where $x^{2}=x^{\mu }x_{\mu }$, and $\mu =0,1,2,3.$ For a scalar field $\psi
(x)$, this transformation law induces a local transformation given by 
\begin{equation}
\psi ^{\prime }(x)=\left[ 1+i\left( \varepsilon ^{\nu }\pi _{\nu }+\frac{1}{2%
}\lambda ^{\mu \nu }L_{\mu \nu }\right) \right] \psi (x),  \label{p4}
\end{equation}%
where $\pi _{\nu }$ is the de Sitter "translation" operator and $\ L_{\mu
\nu }$ is the Lorentz rotation generator, which satisfy the commutation
relations of the de Sitter algebra,%
\begin{align}
\left[ L_{\mu \nu },L_{\rho \sigma }\right] & =\eta _{\mu \sigma }L_{\nu
\rho }-\eta _{\nu \sigma }L_{\mu \rho }-\eta _{\mu \rho }L_{\nu \sigma
}+\eta _{\nu \rho }L_{\mu \sigma }  \notag \\
\left[ L_{\mu \nu },\pi _{\rho }\right] & =\eta _{\nu \rho }\pi _{\mu }-\eta
_{\mu \rho }\pi _{\nu }  \notag \\
\left[ \pi _{\mu },\pi _{\nu }\right] & =-\varepsilon L_{\mu \nu },
\label{p6}
\end{align}%
where $\varepsilon =+1$ corresponds to the algebra $\mathfrak{so(4,1)}$ (See
appendix~\ref{app:ads}).

\subsection{General Relativity in locally-de Sitter spacetimes}

When General Relativity is constructed over a de Sitter background,
spacetime is no longer purely Riemannian and is described by a Cartan
geometry.

Under local de Sitter transformations, the total covariantly conserved
source is the unified current $\Pi ^{\mu \alpha }$ \footnote{%
This tensorial current will reappear in Sec. 4, once the first-order
(vierbein) formalism is introduced, as the Lorentz-frame $3$-form $^{\ast }\Pi _{a}$, the same physical current expressed in the local orthonormal
frame rather than on the coordinate manifold. Following the usual
coordinate-to-frame conversion, the two representations are related by $\Pi
_{ab}=e_{a\mu }e_{b\nu }\Pi ^{\mu \nu }$, with $\Pi _{a}=\Pi _{ab}e^{b}$ the
corresponding frame $1$-form and $^{\ast }\Pi _{a}$ its Hodge dual; see Eq. (%
\ref{9}).}, which couples the symmetric energy-momentum tensor $T^{\mu \nu }$
to the proper conformal current $K^{\mu \alpha }$ \cite{60}-\cite{71} (See
appendix~\ref{app:diffeo}). 
\begin{equation}
\Pi ^{\mu \alpha }=T^{\mu \nu }\xi _{\nu }^{\alpha }=T^{\mu \alpha }-\frac{1%
}{4l^{2}}K^{\mu \alpha },  \label{p7'}
\end{equation}%
where 
\begin{equation}
\xi _{\rho }^{\mu }=\delta _{\rho }^{\mu }-\frac{1}{4l^{2}}\vartheta _{\rho
}^{\mu }\text{ \ and }K^{\mu \alpha }=T^{\mu \nu }\vartheta _{\nu }^{\alpha
},  \label{p8'}
\end{equation}%
are the Killing vectors of the local de Sitter translations, with $\vartheta
_{\rho }^{\mu }=2\eta _{\rho \nu }x^{\nu }x^{\mu }-x^{2}\delta _{\rho }^{\mu
}$ and $x^{2}=\eta _{\mu \nu }x^{\mu }x^{\nu }.$

The modified Einstein field equations that preserve this current are \cite%
{60}-\cite{71} 
\begin{equation}
\mathcal{G}^{\mu \alpha }\equiv \mathcal{R}^{\mu \alpha }-\frac{1}{2}g^{\mu
\alpha }\mathcal{R}=\frac{8\pi G}{c^{4}}\Pi ^{\mu \alpha },  \label{p9'}
\end{equation}%
which splits into the dynamic and kinematic sectors,%
\begin{equation}
G_{(T)}^{\mu \alpha }-G_{(K)}^{\mu \alpha }=\frac{8\pi G}{c^{4}}\left(
T^{\mu \alpha }-\frac{1}{4l^{2}}K^{\mu \alpha }\right) ,  \label{p10'}
\end{equation}%
where $\mathcal{R}^{\mu \alpha }=R_{(T)}^{\mu \alpha }+R_{(K)}^{\mu \alpha }$%
. \ From (\ref{p10'}) it follows that as $l\longrightarrow \infty $, one
recovers Einstein's equations, describing the dynamics of spacetime
curvature. On the other hand, as $l\longrightarrow 0$, we obtain \cite{60}-%
\cite{71},

\begin{equation}
G_{(K)}^{\mu\alpha}=\frac{8\pi G}{4l^{2}c^{4}}K^{\mu\alpha},
\end{equation}
an equation describing the de Sitter effects on the geometry of spacetime. \
In the metric framework of Refs.~\cite{60}-\cite{71}, this shows that the
source of the local cosmological parameter is not the standard mass-energy
current, but the trace of the proper conformal current of matter (see
appendix~\ref{app:diffeo}). In the first-order framework developed below, $%
l(x)$ obeys instead its own field equation, Eq.~(\ref{eq:leom}), and $K^{\mu
\alpha }$ enters only as a source in the vierbein equation (Sec.~\ref{sec:4}%
).

As noted in Sec.~\ref{sec:intro}, the second Bianchi identity forces $%
\Lambda $ to be a global constant in standard General Relativity, which
restricts its use in describing an evolving Universe. The
tensorial de Sitter formalism developed above admits a different consistency
condition, Eq.~(\ref{eq:B.5}) (Appendix~\ref{app:diffeo}), which does not
impose this restriction. Because the source appearing in Eq.~(\ref{p10'}) is the proper conformal current $K^{\mu \alpha }$ rather than the mass-energy current $T^{\mu \nu }$ alone, its local conservation is
compatible with a spacetime-dependent $\Lambda (x)$. This is a direct
consequence of adopting de Sitter-invariant, rather than Poincar\'{e}%
-invariant, local kinematics, as established in Refs.~\cite{60}-\cite{71}.
This observation motivates promoting the de Sitter length scale to a spacetime-dependent field, $l\rightarrow l(x)$.

\section{First-order gauge framework for de Sitter general relativity}

Let $A$ be the de Sitter gauge connection $1$-form taking values in the $%
\mathfrak{so(4,1)}$ Lie algebra. This connection combines the Lorentz spin
connection $\omega ^{ab}$ and the vierbein $1$-form $e^{a}$ into a single
algebraic object, (See appendix~\ref{app:ads}),%
\begin{equation}
A=\frac{1}{2}\omega ^{ab}L_{ab}+\frac{1}{l}e^{a}\pi _{a},  \label{1}
\end{equation}%
where $e^{a}=e_{\text{ \ }\mu }^{a}dx^{\mu }$ is the vierbein 
$1$-form , $\omega ^{ab}=\omega _{\text{ \ \ }\mu }^{ab}dx^{\mu }$ is the
Lorentz connection $1$-form, and $l$ is the (constant) de Sitter length parameter. The $L_{ab}=-L_{ba}$ are the six generators of the Lorentz subalgebra 
$\mathfrak{so(3,1)}$, and $\pi _{a}$ are the four de Sitter "translation"
generators.

The curvature $2$-form $F=dA+A\wedge A$ of the tangent space associated with
the de Sitter connection is obtained by applying the exterior derivative and
the algebra $\mathfrak{so(4,1)}$:%
\begin{eqnarray}
F &=&\frac{1}{2}F^{ab}L_{ab}+F^{a}\pi _{a}  \label{3} \\
&=&\frac{1}{2}\left( R^{ab}-\frac{1}{l^{2}}e^{a}\wedge e^{b}\right) L_{ab}+%
\frac{1}{l}\left( de^{a}+\omega _{b}^{a}\wedge e^{b}\right) \pi _{a},
\label{3a}
\end{eqnarray}%
where $F^{ab}=R^{ab}-\frac{1}{l^{2}}e^{a}\wedge e^{b}$ with $R^{ab}=d\omega
^{ab}+\omega _{c}^{a}\wedge \omega ^{cb}$ the standard Riemann curvature $2$%
-form and $F^{a}=T^{a}/l$ with $T^{a}=de^{a}+\omega _{b}^{a}\wedge e^{b}$
the standard torsion $2$-form.

\subsection{Lanczos-Lovelock action}

The Lanczos-Lovelock (\textbf{LL}) action \cite{76}, \cite{77} is a
polynomial of degree $[d/2]$ in curvature\footnote{%
Here $[x]$ is the integer part of $x$.}, which can also be written in terms
of the Riemann curvature $R^{ab}=d\omega ^{ab}+\omega _{c}^{a}\omega ^{cb}$
and the vielbein $e^{a}$ as\footnote{%
Wedge product between forms is understood throughout.} 
\begin{equation}
S_{G}=\kappa \int \sum_{p=0}^{[d/2]}\alpha _{p}L^{(p)},  \label{Lovaction}
\end{equation}%
where $\alpha _{p}$ are arbitrary constants, and $L^{(p)}$ is given by 
\begin{equation}
L^{(p)}=\epsilon _{a_{1}\cdots a_{d}}R^{a_{1}a_{2}}\!\cdot \!\cdot \!\cdot
\!R^{a_{2p-1}a_{2p}}e^{a_{2p+1}}\!\cdot \!\cdot \!\cdot \!e^{a_{d}}.
\label{Lovlag}
\end{equation}%
In first order formalism the action (\ref{Lovaction}) is regarded as a
functional of the vielbein and the spin connection and the corresponding
field equations obtained varying with respect to $e^{a}$ and $\omega ^{ab}$
read, \cite{80}, \cite{81} 
\begin{equation}
\sum_{p=0}^{[\frac{d-1}{2}]}\alpha _{p}(d-2p)\mathcal{E}_{a}^{p}=0,\text{ \
\ \ \ }\sum_{p=1}^{[\frac{d-1}{2}]}\alpha _{p}p(d-2p)\mathcal{E}_{ab}^{p}=0,
\label{E-L}
\end{equation}%
where \cite{80}, \cite{81} 
\begin{eqnarray*}
\mathcal{E}_{a}^{p}:= &&\epsilon _{ab_{1}\cdots b_{d-1}}R^{b_{1}b_{2}}\cdots
R^{b_{2p-1}b_{2p}}e^{b_{2p+1}}\cdots e^{b_{d-1}},\hspace{-0.06in} \\
\mathcal{E}_{ab}^{p}:= &&\epsilon _{aba_{3}\cdots a_{d}}R^{a_{3}a_{4}}\cdots
R^{a_{2p-1}a_{2p}}T^{a_{2p+1}}e^{a_{2p+2}}\cdots e^{a_{d}},\hspace{-0.06in}
\end{eqnarray*}%
with $T^{a}=de^{a}+\omega _{b}^{a}e^{b}$ is the torsion $2$-form.

The first two terms in the LL action (\ref{Lovaction}) are the cosmological
and kinetic terms of the Einstein Hilbert Cartan action (\ref{cri3}), and
therefore General Relativity is contained in the LL theory as a particular
case.

For a given dimension and an arbitrary choice of coefficients $\alpha _{p}$%
's, higher dimensional LL theories have some drawbacks. One difficulty can
be viewed in the static, spherically symmetric solutions of (\ref{E-L}). For
arbitrary $\alpha _{p}$'s there are negative energy solutions with horizons
and positive energy solutions with naked singularities \cite{82}.

These problems can be curbed if the coefficients $\alpha _{p}$'s are chosen
in a suitable way. In Ref. \cite{81} was shown that requiring the theories
to possess \emph{a unique} cosmological constant, strongly restricts the
coefficients $\alpha _{p}$'s. As a consequence, one obtains a set of
theories labelled by an integer $k$ which lead to well defined black hole
configurations. These theories are described by the action 
\begin{equation}
S_{k}=\kappa \int \sum_{p=0}^{k}c_{p}^{k}L^{(p)}\;,  \label{Ik}
\end{equation}%
which is obtained from (\ref{Lovaction}) with the choice 
\begin{equation}
\alpha _{p}:=c_{p}^{k}=\left\{ 
\begin{array}{ll}
\frac{l^{2(p-k)}}{(d-2p)}\left( 
\begin{array}{c}
k \\ 
p%
\end{array}%
\right) & ,\text{ }p\leq k \\ 
0 & ,\text{ }p>k%
\end{array}%
\right.  \label{Coefs}
\end{equation}%
where $1\leq k\leq \lbrack \frac{d-1}{2}]$.

For a given dimension $d$, the coefficients $c_{p}^{k}$ give rise to a
family of inequivalent theories, labeled by the integer $k\in \{1,...,[\frac{%
d-1}{2}]\}$ which represents the highest power of curvature in the
Lagrangian. This set of theories possesses only two fundamental
constants, $\kappa $ and $l$, related to the gravitational constant $G_{k}$
and the cosmological constant $\Lambda $ through 
\begin{eqnarray}
\kappa &=&\frac{1}{2(d-2)!\Omega _{d-2}G_{k}},  \label{Kappa} \\
\Lambda &=&\pm \frac{(d-1)(d-2)}{2l^{2}}.  \label{Lambda}
\end{eqnarray}

\subsection{The Einstein-Hilbert-Cartan action}

The Einstein-Hilbert-Cartan action with cosmological constant in $d$
dimensions is recovered setting $k=1$ in (\ref{Ik}) using (\ref{Coefs}) and (%
\ref{Kappa}, \ref{Lambda}). In fact, for $k=1$, Eq.~(\ref{Ik}) takes the
form 
\begin{equation}
S_{1}=S_{EHC}=\kappa \int \left\{ c_{0}^{1}L^{(0)}+c_{1}^{1}L^{(1)}\right\}
\label{cri2}
\end{equation}%
where, $L^{(0)}$ and $L^{(1)}$ for the $d=4$ case can be obtained from (\ref%
{Lovlag}), 
\begin{eqnarray}
L^{(0)} &=&\epsilon _{abcd}\,e^{a}\wedge e^{b}\wedge e^{c}\wedge e^{d}
\label{L0d4} \\
L^{(1)} &=&\epsilon _{abcd}\,R^{ab}\wedge e^{c}\wedge e^{d},\text{ }
\label{L1d4}
\end{eqnarray}%
and from (\ref{Coefs}), we find that for $d=4$, $c_{0}^{1}=1/4\,l^{2},$ $%
c_{1}^{1}=1/2.$ So that, 
\begin{equation}
S_{EHC}=\frac{\kappa }{2}\int \varepsilon _{abcd}R^{ab}e^{c}\wedge e^{d}\mp 
\frac{1}{2\,l^{2}}\varepsilon _{abcd}e^{a}\wedge e^{b}\wedge e^{c}\wedge
e^{d},  \label{cri3}
\end{equation}%
with $\left. \kappa \right\vert _{d=4}=1/16\pi G$ and $\Lambda =\pm 3/l^{2},$
which is the Einstein-Hilbert action with cosmological constant ($c=1$).

\subsection{Modified Einstein-Hilbert-Cartan action}

\label{sec:desitter}

If the local kinematics of the Universe is governed by the de Sitter group $%
SO(4,1)$, the local tangent space is no longer flat and acquires an
intrinsic kinematical curvature governed by a length parameter $l(x)$. To
formulate this without the ambiguities of the standard metric framework, we
promote the fixed, rigid scale $l$ of Eqs.~(\ref{1})-(\ref{3a}) to a
spacetime field, $l\rightarrow l(x)$.

Let $A$ be the de Sitter gauge connection $1$-form taking values in the $%
\mathfrak{so(4,1)}$ Lie algebra. This connection combines the Lorentz spin
connection $\omega ^{ab}$ and the vierbein $1$-form $e^{a}$ into a single
algebraic object, (See appendix~\ref{app:ads}),%
\begin{equation}
A=\frac{1}{2}\omega ^{ab}L_{ab}+\frac{1}{l(x)}e^{a}\pi _{a},
\end{equation}%
where $e^{a}=e_{\text{ \ }\mu }^{a}dx^{\mu }$ is the vierbein $1$-form , $%
\omega ^{ab}=\omega _{\text{ \ \ }\mu }^{ab}dx^{\mu }$ is the Lorentz
connection $1$-form, $l(x)$ is the local length parameter, $L_{ab}=-L_{ba}$
are the six generators of the Lorentz subalgebra $\mathfrak{so(3,1)}$, and $%
\pi _{a}$ are the four de Sitter "translation" generators.

The curvature components $F=\frac{1}{2}F^{ab}L_{ab}+F^{a}\pi _{a}$ take now
the form 
\begin{align}
F^{ab}& =R^{ab}-\frac{1}{l^{2}(x)}e^{a}\wedge e^{b}  \label{4} \\
F^{a}& =\frac{1}{l(x)}\left[ T^{a}-\frac{1}{l(x)}dl(x)\wedge e^{a}\right] ,
\label{4a}
\end{align}%
with $T^{a}=de^{a}+\omega _{b}^{a}\wedge e^{b}$ the standard torsion $2$%
-form.

To formulate de Sitter General Relativity in the first-order formalism, we
propose a gravitational action in four dimensions given by%
\begin{equation}
S=S_{g}+S_{m}=S_{MEHC}+S_{NY}+S_{m}
\end{equation}%
where $S_{g}=S_{MEHC}+S_{NY}$ is the gravitational action, $S_{MEHC}$ is now
the four-dimensional modified Einstein-Hilbert action with cosmological
constant, given by 
\begin{equation}
S_{MEHC}=\frac{\kappa }{2}\int \varepsilon _{abcd}R^{ab}e^{c}\wedge e^{d}-%
\frac{\kappa }{4}\int \frac{1}{l^{2}(x)}\varepsilon _{abcd}e^{a}\wedge
e^{b}\wedge e^{c}\wedge e^{d},  \label{6}
\end{equation}%
$S_{NY}$ is the Nieh-Yan term, and $S_{m}$ corresponds to the action for
matter fields.

In this first-order gauge approach, the cosmological term is not appended as
a separate correction to the Einstein-Hilbert action; it arises as a
component of the $\mathfrak{so(4,1)}$ curvature in the tangent space. This
provides the basis for deriving the field equations, conservation laws, and
cosmological consequences discussed in the following sections.

\subsection{The Nieh-Yan term $S_{NY}$}

\label{sec:nieh-yan-eom}

Before introducing the Nieh-Yan term, note that $l(x)$ enters $S_{MEHC}$,
Eq. (\ref{6}), only algebraically, through the volume term: no derivative of 
$l(x)$ appears anywhere in $S_{MEHC}$. Consequently, varying $S_{MEHC}$ with
respect to $l(x)$ alone forces 
\begin{equation}
\frac{\kappa }{2l^{3}(x)}\varepsilon _{abcd}e^{a}\wedge e^{b}\wedge
e^{c}\wedge e^{d}=0,  \label{eq:degenerate}
\end{equation}%
i.e., a degenerate spacetime volume form for any nondegenerate vierbein: $%
l(x)$, as it stands in Eq.~(\ref{6}), admits no independent equation of
motion, and cannot consistently be treated as a dynamical field. The
Nieh-Yan term \cite{74} is given by the following topological action 
\begin{equation}
S_{NY}=-2\alpha \kappa \int dl(x)\wedge T^{a}\wedge e_{a},  \label{eq:SNY}
\end{equation}%
where $\alpha $ is a coupling constant with dimensions of inverse length,
normalized with the same $\kappa $ scale as $S_{MEHC}$ so that it enters on
the same footing as the gravitational sector. The overall sign is a free convention, since $\alpha $ is odd under P and T (Sec.~\ref{sec:symm}). Moreover, in the absence of spin sources, the field equations for the
geometry and for $l(x)$ depend on $\alpha $ only through $\alpha ^{2}$ [Eq.~(%
\ref{eq:ldampedosc}) and Appendix~\ref{app:Seff}]. Only the relative sign
between $\alpha $ and an axial spin density is physical. This follows a
Peccei-Quinn-type mechanism used for an axion-like pseudoscalar field by
Mercuri \cite{83} and by Taveras and Yunes \cite{84}, applied here to $l(x)$
itself. A gauge-theoretic origin for this coupling is suggested by the
Pontryagin density of the $\mathfrak{so}(4,1)$ curvature. With Eq.~(\ref%
{ds13}), 
\begin{equation*}
F_{AB}\wedge F^{AB}=R_{ab}\wedge R^{ab}+\frac{2}{l^{2}}\left( T^{a}\wedge
T_{a}-e^{a}\wedge e^{b}\wedge R_{ab}\right) ,
\end{equation*}%
so that the Nieh-Yan density appears with the coefficient $2/l^{2}$, in line
with the proportionality between the Nieh-Yan coupling and the cosmological
constant found in pre-geometric gravity \cite{85,86}. A closely related construction has been given in Ref.~\cite{87}. There, gravity arises as a deformation of a topological gauge theory on the Cartan geometries $SO(4,1)/SO(3,1)$ and $SO(3,2)/SO(3,1)$, and a Nieh-Yan-type term appears with couplings fixed by the bare cosmological constant.
Its extension to the M\"{o}bius group introduces a dilation scalar with its
own kinetic term, coupled to torsion. For constant $l$ this term is a total
derivative, but for $l=l(x)$ an integration by parts gives $\int
(2/l^{2})\,d(e^{a}\wedge T_{a})=\int (4/l^{3})\,dl\wedge e^{a}\wedge T_{a}$
up to a boundary term, which has the form of Eq.~(\ref{eq:SNY}) with a
field-dependent coupling $\propto 1/l^{3}$. In this work we keep $\alpha $
constant. The consequences of this addition, an independent field equation
for $l(x)$, and its associated corrections to the torsion and vierbein
equations, are derived in Sec.~\ref{sec:4} and Sec.~\ref{sec:5}.

\subsection{First-Order Action for Matter Fields}

We now consider how matter fields couple to this de Sitter geometry. \ In
the first-order formalism, matter is minimally coupled to the vierbein $%
e^{a} $ and the Lorentz connection $\omega^{ab}$. Requiring invariance of
the matter action under local de Sitter transformations, the generalized
Noether theorem yields a unified conserved current that couples the standard
energy-momentum tensor with the proper conformal current of matter.

In the first-order Palatini formalism, the action for a matter field $\psi$
in a four-dimensional region $\Omega$ is

\begin{equation}
S_{m}=\int_{\Omega }\mathcal{L}_{m}(e^{a},\omega ^{ab},\psi ),  \label{7}
\end{equation}%
where $\mathcal{L}_{m}$ is the matter Lagrangian density $4$-form and $%
e=\det \left( e_{\mu }^{a}\right) =\sqrt{-g}$ is the volume measure. Unlike
the metric formulation, where the matter Lagrangian depends only on $g_{\mu
\nu }$, here the vierbein $e^{a}$ and the Lorentz connection $\omega ^{ab}$
are independent gauge fields. Varying the matter action with respect to
these independent fields defines two current densities:

$\left( i\right) $ The energy-momentum tensor $\mathbf{\tau}_{a}^{\mu}$,
representing the response of matter fields to local translations in the
tangent space:%
\begin{equation}
\mathbf{\tau}_{a}^{\mu}=\frac{1}{e}\frac{\delta\mathcal{L}_{m}}{\delta
e_{\mu }^{a}},  \label{7a}
\end{equation}
which, in the absence of intrinsic spin, relates to the symmetric metric
energy-momentum tensor via $T^{\mu\nu}=\mathbf{\tau}_{a}^{\mu }e^{a\nu}$.

$\left( ii\right) $ The tangent spin-density tensor $\mathcal{S}%
_{[ab]}^{\mu} $, measuring the intrinsic angular momentum and spin density
of matter fields, acting as the source of local Lorentz rotations:%
\begin{equation}
\mathcal{S}_{[ab]}^{\mu}=\frac{1}{e}\frac{\delta\mathcal{L}_{m}}{\delta
\omega_{\mu}^{ab}}.  \label{7b}
\end{equation}

\subsection{Gauge Invariance and Consistency of the Localized Scale Action}

\label{sec:symm}

A relevant point when introducing a spacetime-dependent length parameter $%
l(x)$\ into the action concerns the status of the local $\mathfrak{so}(4,1)$%
\ gauge invariance. In standard Lanczos-Lovelock gravity in $4D$\ \cite%
{76,77}, the de Sitter radius $l_{0}$\ is a fixed constant, and the
connection

\begin{equation}
\hat{A}=\frac{1}{2}\omega ^{ab}L_{ab}+\frac{1}{l_{0}}\hat{e}^{a}\pi _{a},
\label{1a}
\end{equation}%
takes values in the de Sitter algebra $\mathfrak{so}(4,1)$: under a local
translation parameter $\zeta ^{a}$, $\hat{e}^{a}$\ shifts as $\delta \hat{e}%
^{a}=D\zeta ^{a}$, while $\omega ^{ab}$ shifts by a term proportional to $%
l_{0}^{-2}(\zeta ^{a}\hat{e}^{b}-\zeta ^{b}\hat{e}^{a})$. The rigid action,
built with the same Lanczos-Lovelock structure as Eq.~(\ref{6}) but with the
fixed scale\textbf{\ }$l_{0}$\textbf{,} 
\begin{equation}
\hat{S}_{g}=\frac{\kappa }{2}\int \varepsilon _{abcd}R^{ab}\wedge \hat{e}%
^{c}\wedge \hat{e}^{d}-\frac{\kappa }{4}\int \frac{1}{l_{0}^{2}}\varepsilon
_{abcd}\hat{e}^{a}\wedge \hat{e}^{b}\wedge \hat{e}^{c}\wedge \hat{e}^{d},
\label{1a'}
\end{equation}%
is invariant under local Lorentz transformations, but not
under these local translations. As in first-order Einstein-Hilbert gravity with a cosmological constant, these translations are symmetries only on the torsion-free shell, where they coincide with diffeomorphisms.

To relate this rigid construction with the localized-scale framework of Sec.~%
\ref{sec:desitter}, we connect the physical vierbein $e^{a}$\ to the
reference field $\hat{e}^{a}$\ through a local Weyl-type rescaling by the
dimensionless ratio\textbf{\ }$l(x)/l_{0}$\textbf{:}%
\begin{equation}
e^{a}=\frac{l(x)}{l_{0}}\hat{e}^{a}.  \label{1b}
\end{equation}

\textbf{Curvature sector}. Since $R^{ab}$\ depends only on $\omega ^{ab}$,
it is unaffected by $Eq.$\ (\ref{1b}). Direct substitution gives\textbf{\ }%
\begin{equation}
\hat{e}^{a}\wedge \hat{e}^{b}=\left( \frac{l_{0}}{l(x)}\right)
^{2}e^{a}\wedge e^{b},  \label{1c}
\end{equation}%
so that%
\begin{equation}
\hat{F}^{ab}=R^{ab}-\frac{1}{l_{0}^{2}}\hat{e}^{a}\wedge \hat{e}^{b}=R^{ab}-%
\frac{1}{l^{2}(x)}e^{a}\wedge e^{b},  \label{1d}
\end{equation}%
which reproduces $Eq.$(\ref{4}) identically.

\textbf{Torsional sector.} Applying the exterior derivative to $Eq.$ (\ref%
{1b}),%
\begin{equation}
\hat{T}^{a}=d\hat{e}^{a}+\omega_{b}^{a}\hat{e}^{b}=\frac{l_{0}}{l(x)}\left(
de^{a}+\omega_{b}^{a}e^{b}\right) +l_{0}d\left( \frac{1}{l(x)}\right) \wedge
e^{a},  \label{2a}
\end{equation}
so that%
\begin{equation}
\frac{1}{l_{0}}\hat{T}^{a}=\frac{1}{l(x)}T^{a}-\frac{1}{l^{2}(x)}dl(x)\wedge
e^{a}=F^{a},  \label{2b}
\end{equation}
reproducing $Eq$. (\ref{4a}) identically. This confirms that the gradient
term $dl(x)$ appearing in $F^{a}$ is the exact Jacobian of the local Weyl
rescaling (\ref{1b}).

A direct calculation shows that Eqs.~(\ref{4})-(\ref{4a}) are the pullback
of the rigid, covariant curvature $\hat{F}^{AB}$\ under this rescaling,
while the action itself is not invariant: $\hat{S}_{g}\neq S_{g}$, differing
by a local conformal weight. Substituting Eq.~(\ref{1c}) and its
four-vierbein analogue $\hat{e}^{a}\hat{e}^{b}\hat{e}^{c}\hat{e}%
^{d}=(l_{0}/l(x))^{4}e^{a}e^{b}e^{c}e^{d}$ into the integrand of Eq.~(\ref%
{1a'}), both terms factor the same overall weight, giving the exact relation
between the Lagrangian $4$-forms 
\begin{equation}
\hat{\mathcal{L}}_{g}=\left( \frac{l_{0}}{l(x)}\right) ^{2}\left[ \frac{%
\kappa }{2}\varepsilon _{abcd}R^{ab}\wedge e^{c}\wedge e^{d}-\frac{\kappa }{4%
}\frac{1}{l^{2}(x)}\varepsilon _{abcd}e^{a}\wedge e^{b}\wedge e^{c}\wedge
e^{d}\right] =\left( \frac{l_{0}}{l(x)}\right) ^{2}\mathcal{L}_{MEHC}.
\label{2c}
\end{equation}%
Since the weight $(l_{0}/l(x))^{2}$ depends on the point, it cannot be taken
outside the integral: the two actions $\hat{S}_{g}=\int \hat{\mathcal{L}}%
_{g} $ and $S_{MEHC}=\int \mathcal{L}_{MEHC}$ are different functionals. The
Nieh-Yan term carries the same weight: since $d(l_{0}/l)$ is proportional to 
$dl$, one finds $dl\wedge \hat{T}^{a}\wedge \hat{e}_{a}=(l_{0}/l(x))^{2}\,dl%
\wedge T^{a}\wedge e_{a}$, so the extension by $S_{NY}$ does not alter this
conclusion.

This is a feature of the construction. It shows that the
localized-scale action~(\ref{6}) is not the direct pullback of a rigid $%
SO(4,1)$-invariant functional, and that full $SO(4,1)$\ covariance of $S_{g}$%
\ under the complete de Sitter group, including translations $\pi _{a}$,
cannot be claimed without qualification.

\textbf{Symmetry breaking }$SO(4,1)\longrightarrow SO(3,1)$\textbf{. }The rigid, fixed-$l_{0}$\ construction of Eqs.~(\ref{1a})-(\ref{1a'}%
) is already understood, following Refs.~\cite{77,80,81}, as a
formally $SO(4,1)$-covariant connection whose Lagrangian is built entirely
from $SO(3,1)$\ tensors. The action is invariant under the local Lorentz
subgroup $SO(3,1)$\ acting on $\omega ^{ab}$\ and $e^{a}$\ at fixed $l(x)$. Both terms of $\hat{S}_{g}$ in Eq.~(\ref{1a'}), $\varepsilon
_{abcd}R^{ab}\wedge \hat{e}^{c}\wedge \hat{e}^{d}$ and $\varepsilon _{abcd}%
\hat{e}^{a}\wedge \hat{e}^{b}\wedge \hat{e}^{c}\wedge \hat{e}^{d}$, are built from $SO(3,1)$\ tensors ($R^{ab}$ a Lorentz curvature, $\hat{e}^{a}$ a Lorentz vector). However, neither term is invariant under the full local translational sector of $SO(4,1)$. This breaking is explicit
rather than spontaneous. It is built into the structure of the action itself and does not arise from a dynamically selected vacuum of an otherwise $SO(4,1)$-symmetric Lagrangian. In this restricted sense, it is already present in the fixed-$l_{0}$\ case. What is
qualitatively new when $l\rightarrow l(x)$\ is not the existence of this
breaking per se, but its promotion from a fixed, spacetime-independent
choice to a local, dynamical field, governed by the field equations of Secs.~%
\ref{sec:4}-\ref{sec:5}. The field $l(x)$\ can be regarded as a compensator field,
loosely analogous to a Higgs field in symmetry-breaking gauge theories of gravity \cite{75,85,86}. Unlike a Higgs field, however, it has no potential $V(l)$\ in the action and no vacuum degeneracy. An effective potential $V\propto 1/l^{2}$ emerges only after the torsion is eliminated (Sec.~\ref{sec:6.2}). The gradient $dl(x)$\ does not measure the breaking, which is already present when $dl=0$ in the fixed-$l_{0}$\ case. It measures instead the local departure of this
compensator field from the fixed reference value $l_{0}$, entering the
dynamics through the field equations of Secs.~\ref{sec:5}-\ref{sec:6}.
Consistency of the theory's gauge field equations requires matter spin
density to source torsion, with the axial (chiral) component of that same
torsion, in turn, sourcing $dl(x)$\ through the Nieh-Yan invariant, as
derived in Secs.~\ref{sec:5}-\ref{sec:6}. This is consistent with the result of Ref.~\cite{73} for the same class of MacDowell-Mansouri-type actions. Such constructions admit a description equivalent to General Relativity via the Weyl rescaling $g_{\mu \nu }\rightarrow f^{2}g_{\mu \nu }$, $l\rightarrow fl$. In this description, $l$\ is absorbed into the metric away from points where the rescaling degenerates. The present work does not carry out this absorption explicitly. Instead,
Sec.~\ref{sec:4.2}\ verifies directly that the form-language field equation~(%
\ref{ecmov}) is equivalent to its tensorial counterpart with $\Lambda (x)=3/l^{2}(x)$. The torsion sector derived in Secs.~\ref{sec:5}-\ref{sec:6} lies outside the purely metric formulation of Ref.~\cite{73}. For a systematic account of symmetry breaking and dynamics in de
Sitter gauge gravity, see Ref.~\cite{75}.

\textbf{Symmetry group of the complete action.} The complete action $%
S=S_{MEHC}+S_{NY}+S_{m}$ is invariant under 
\begin{equation}
\mathrm{Diff}(M)\ltimes SO(3,1)_{\mathrm{local}},  \label{eq:symgroup}
\end{equation}%
the same gauge group as Einstein-Cartan theory, and under no larger group of
local transformations. This statement refers to the field content $%
(e^{a},\omega ^{ab},l)$ used here. Appendix~\ref{app:cartan} shows that, by
introducing the Stelle-West field $y^{A}$, of which these fields are the
physical-gauge components, the same action can be rewritten in a manifestly $SO(4,1)$-invariant form, with the breaking appearing as the
gauge choice $y^{A}\propto \delta _{4}^{A}$. The two descriptions are
equivalent: the second is a nonlinear (St\"{u}ckelberg-type) realization
that adds no physical degrees of freedom. Invariance under diffeomorphisms
follows because every term is the integral of a $4$-form built without
background structures. Local Lorentz invariance follows because every term
is a complete contraction of Lorentz tensors ($e^{a}$, $T^{a}$, $R^{ab}$, $%
\eta _{ab}$, $\varepsilon _{abcd}$) with the scalar $l(x)$. The Nieh-Yan
term does not reduce this group: $dl\wedge T^{a}\wedge e_{a}$ is the product
of the scalar $1$-form $dl$ with the Lorentz-invariant $3$-form $T^{a}\wedge
e_{a}$. The breaking $SO(4,1)\rightarrow SO(3,1)$ is therefore already
present in $S_{MEHC}$, as discussed above; $S_{NY}$, which is not invariant
under the translations either, neither reduces the Lorentz group nor
restores the translational sector. The six local de Sitter
translations generated by $\pi _{a}$ are not symmetries of the action. For
constant $l$ they map solutions into solutions only when $T^{a}=0$, where
they coincide on shell with diffeomorphisms. Once $l=l(x)$ and the Nieh-Yan
term are included, the torsion no longer vanishes on shell, Eq.~(\ref{eq:D8}%
), and not even this on-shell correspondence survives.

The de Sitter group survives in two weaker senses. First, as the algebra in
which the Cartan connection takes values, Eq.~(\ref{1a}), a geometric rather
than a dynamical statement. Second, as the isometry group of the vacuum, but
only approximately: for $\alpha \neq 0$, de Sitter space with constant $l$
and $T^{a}=0$ is not an exact solution, since Eq.~(\ref{eq:leom}) with $%
\mathcal{N}=0$ requires $l\rightarrow \infty $. The ten de Sitter isometries
are realized only approximately along the slow-roll attractor of Sec.~\ref%
{sec:6.2}, where $H\simeq 1/l$, and are broken by the rolling of $l(t)$, as
for any quintessence background. Asymptotically $l\rightarrow \infty $ and $%
\Lambda \rightarrow 0$, and the vacuum approaches Minkowski space, with
Poincar\'{e} symmetry.

\textbf{Discrete symmetries.} The Nieh-Yan density $\mathcal{N}$ is a
pseudoscalar, odd under parity and under time reversal, whereas $l(x)$,
being a positive length, is a true scalar. The coupling $S_{NY}$ therefore
breaks P and T explicitly, while preserving PT, and hence remains compatible
with CPT. Equivalently, the coupling constant $\alpha $ is odd under P and
under T. In the absence of matter coupled to the connection, the effective
action~(\ref{eq:D10}) depends on $\alpha $ only through $\alpha ^{2}$ and is
parity-even, so this violation does not show up in the gravitational sector
or in the cosmological background. It appears in the coupling to fermions:
eliminating the torsion in the presence of Dirac fermions generates the term 
$\alpha \,\partial _{\mu }l\,\bar{\psi}\gamma ^{5}\gamma ^{\mu }\psi $, odd
under P and T, analogous to the derivative coupling of an axion (Appendix~%
\ref{app:Seff}). This is where the parity violation induced by the Nieh-Yan
coupling would be observable.

The field equations of Sec.~\ref{sec:4}, and the torsion dynamics of Secs.~%
\ref{sec:5}-\ref{sec:6}, all follow from Eqs.~(\ref{4})-(\ref{4a})
independently of which reading is adopted for the global symmetry status of
the action.

\section{Field Equations and Equivalence with the Tensor Formalism}

\label{sec:4}

By establishing the action functional of the modified
Einstein-Hilbert-Cartan gauge framework and characterizing the material
sources, the next step is to obtain the complete set of gravitational field
equations. Within the first-order Palatini approach, the vierbein $e^{a}$
and the Lorentz connection $\omega^{ab},$ and the length field $l(x)$ are
treated as independent variables, each yielding its own field equation.

\subsection{Variation with respect to $l(x)$}

\label{sec:4.1}

We consider first the variation of the total action $S_{MEHC}+S_{NY}+S_{m}$
with respect to $l(x),$ holding $e^{a}$ and $\omega ^{ab}$ fixed.
Integrating the Nieh-Yan term by parts (the boundary term vanishes for fixed
boundary conditions), we obtain 
\begin{equation}
\delta _{l}S_{NY}=2\alpha \kappa \int \delta l\;d\left( T^{a}\wedge
e_{a}\right) =2\alpha \kappa \int \delta l\;\mathcal{N}(x)\,e\,d^{4}x,
\label{eq:dlSNY}
\end{equation}%
Using the structure equations $dT^{a}=R^{a}{}_{b}\wedge e^{b}-\omega
^{a}{}_{b}\wedge T^{b}$ and $de_{a}=T_{a}-\omega _{a}{}^{b}\wedge e_{b}$,
the connection-dependent terms cancel identically, leaving the Nieh-Yan
4-form, 
\begin{equation}
d\left( T^{a}\wedge e_{a}\right) =T^{a}\wedge T_{a}-e^{a}\wedge e^{b}\wedge
R_{ab}\equiv \mathcal{N}.  \label{eq:NY}
\end{equation}
Here and below the same symbol denotes the $4$-form and its component, $%
\mathcal{N}=\mathcal{N}(x)\,e\,d^{4}x$. The variation of the volume term of $%
S_{MEHC}$ gives%
\begin{equation}
\delta _{l}S_{MEHC}=12\kappa \int \frac{\delta l}{l^{3}}\,e\,d^{4}x,
\label{dvol}
\end{equation}%
where we have used $\varepsilon _{abcd}e^{a}e^{b}e^{c}e^{d}=4!\,e\,d^{4}x.$

Combining Eq.~(\ref{eq:dlSNY}) with the variation of the volume term of $%
S_{MEHC}$ [Eq.~(\ref{dvol})], we find that the $l(x)$ equation of motion is 
\begin{equation}
\frac{1}{l^{3}(x)}=-\frac{\alpha }{6}\,\mathcal{N}(x).  \label{eq:leom}
\end{equation}%
As it stands this is algebraic in $l(x)$: no derivatives of $l$ appear, and $%
l$ is fixed pointwise by the local torsion-curvature invariant $\mathcal{N}$%
. However, $\mathcal{N}$ depends on the axial torsion, which the torsion
equation of Sec.~\ref{sec:5} fixes in terms of $\partial l$; once torsion is
eliminated, Eq.~(\ref{eq:leom}) becomes a second-order equation for $l(x)$,
as shown for FLRW in Sec.~\ref{sec:6.2}. As $\alpha \rightarrow 0$, Eq.~(\ref%
{eq:leom}) correctly reduces to Eq.~(\ref{eq:degenerate}), confirming that
the Nieh-Yan term is the missing ingredient identified in Sec.~\ref%
{sec:nieh-yan-eom}.

In components, writing $R_{ab}=\frac{1}{2}R_{abcd}e^{c}\wedge e^{d}$ and $%
T^{a}=\frac{1}{2}T^{a}{}_{bc}e^{b}\wedge e^{c}$, and using $e^{a}\wedge
e^{b}\wedge e^{c}\wedge e^{d}=-\varepsilon ^{abcd}\,e\,d^{4}x$, where $%
\varepsilon ^{abcd}$ is obtained from $\varepsilon _{0123}=+1$ by raising
indices with $\eta _{ab}$, so that $\varepsilon ^{0123}=-1$ (the convention
used throughout this paper), 
\begin{equation}
\mathcal{N}=\mathcal{T}^{2}-\mathcal{R}^{\ast },\qquad \mathcal{T}^{2}=-%
\frac{1}{4}\varepsilon ^{bcde}T^{a}{}_{bc}T_{a\,de},\qquad \mathcal{R}^{\ast
}=-\frac{1}{2}\varepsilon ^{abcd}R_{abcd}.  \label{eq:NYcomp}
\end{equation}%
The torsion $2$-form decomposes irreducibly into trace, axial and tensor
parts. For torsion of purely trace type, $T^{a}{}_{bc}=\delta
_{b}^{a}V_{c}-\delta _{c}^{a}V_{b}$, i.e.\ $T^{a}=e^{a}\wedge V$ with $%
V=V_{c}e^{c}$, one has $T^{a}\wedge e_{a}=-V\wedge e^{a}\wedge e_{a}=0$
identically, and hence $\mathcal{N}=d(T^{a}\wedge e_{a})=0$; separately, $%
\mathcal{T}^{2}=-\varepsilon ^{bcde}\eta _{bd}V_{c}V_{e}=0$. A non-vanishing
axial (parity-odd) torsion component is therefore required to activate Eq.~(%
\ref{eq:leom}), as made explicit in the cosmological ansatz of Sec.~\ref%
{sec:6}. Note that minimally coupled Dirac fermions source this axial
component, through the axial current $\bar{\psi}\gamma ^{5}\gamma ^{a}\psi $%
, whereas trace torsion requires a spin-fluid or non-minimally coupled
source.

\subsection{Vierbein variation and the de Sitter field equations}

We begin by applying the principle of stationary action with respect to the
vierbein $1$-form $e^{a}$, while keeping the Lorentz spin connection $%
\omega^{ab}$ and the localized parameter $l(x)$ fixed. The total variation
of the action $S=S_{MEHC}+S_{NY}+S_{m}$ must vanish ($\delta_{e}S=0$).

The geometric variation splits into the standard Riemann-Palatini variation
and the kinematic volume variation. One finds 
\begin{equation}
\delta _{e}S_{MEHC}=\int \left[ -\kappa \,\varepsilon
_{abcd}R^{bc}e^{d}+\kappa \,\frac{1}{l^{2}(x)}\varepsilon _{abcd}e^{b}\wedge
e^{c}\wedge e^{d}\right] \wedge \delta e^{a}.  \label{8}
\end{equation}

For the Nieh-Yan term [Eq.~(\ref{eq:SNY})] we find, 
\begin{equation}
\delta_{e}S_{NY}=-4\alpha\kappa\int dl(x)\wedge T_{a}\wedge\delta e^{a}.
\label{eq:deSNY}
\end{equation}

On the other hand, the variation of the matter action with respect to the
vierbein defines the $3$-form dual of the de Sitter energy-momentum current (%
$^{\ast}\Pi_{a}$), which contains both the standard energy-momentum tensor
and the proper conformal current. This is the same de Sitter energy-momentum
current defined in Sec. $2.1$, Eq. (\ref{p7'}), now expressed as a
Lorentz-frame $3$-form: $^{\ast}\Pi_{a}$ is the Hodge dual of the $1$-form $%
\Pi_{a}=\Pi_{ab}e^{b}$, obtained from $\Pi^{\mu\alpha}$ by contraction with
the vierbein, $\Pi_{ab}=e_{a\mu}e_{b\nu}\Pi^{\mu\nu}$. Throughout Secs.~\ref%
{sec:4}-\ref{sec:6} we set $\hbar=c=1$, consistent with the $\hbar$- and $c$%
-independent normalization of $\kappa$ already adopted in Eq.~(\ref{Kappa}); 
$S_{m}=\int_{\Omega}\mathcal{L}_{m}$ in these natural units, so that:

\begin{equation}
\delta _{e}S_{m}=\int_{\Omega }\text{ }^{\ast }\Pi _{a}\wedge \delta
e^{a}=\int_{\Omega }\text{ }\left( ^{\ast }\tau _{a}-\frac{1}{4l^{2}(x)}%
^{\ast }K_{a}\right) \wedge \delta e^{a}.  \label{9}
\end{equation}%
Since $\mathcal{L}_{m}$ in Eq.~(\ref{7}) does not depend on $%
l(x)$, the conformal term in Eq.~(\ref{9}) is not generated by an explicit $%
l $-dependence of the matter Lagrangian. It is the frame form of the de
Sitter Noether current $\Pi ^{\mu \alpha }$ of Eq.~(\ref{p7'}) (Appendix~\ref%
{app:diffeo}), which we identify with the vierbein response of the matter
sector. Consistently, $S_{m}$ does not contribute to the variation with
respect to $l(x)$, Eq.~(\ref{eq:leom}).

Demanding that $\delta _{e}S_{MEHC}+\delta _{e}S_{NY}+\delta _{e}S_{m}=0$
for any arbitrary and independent variation $\delta e^{a}$, we arrive at the de Sitter-Palatini Field Equation in the language of
differential forms:%
\begin{equation}
\frac{1}{2}\varepsilon _{abcd}R^{bc}e^{d}-\frac{1}{2l^{2}(x)}\varepsilon
_{abcd}e^{b}\wedge e^{c}\wedge e^{d}+2\alpha \,dl(x)\wedge T_{a}=8\pi
G\left( ^{\ast }\tau _{a}-\frac{1}{4l^{2}(x)}\text{ }^{\ast }K_{a}\right) .
\label{ecmov}
\end{equation}

This $3$-form equation unifies the entire system: the dynamic Riemann
curvature ($R^{bc}$) couples to the mass-energy source ($^{\ast}\tau_{a}$),
while the kinematic volume of the tangent space ($e^{b}\wedge e^{c}\wedge
e^{d}$) couples directly to the proper conformal current ($^{\ast}K_{a}$).

\subsection{Equivalence with the Tensorial Formalism}

\label{sec:4.2}

To ensure the consistency of the theory, the left-hand side of the $3$-form
gauge equation must map onto the tensorial splitting $G_{(T)}^{\mu \nu
}-G_{(K)}^{\mu \nu }$ developed in the coordinate manifold in \cite{60}-\cite%
{71}. We convert every $3$-form in Eq.~(\ref{ecmov}) with one and the same
rule. Any $3$-form $\Omega =\frac{1}{3!}\Omega _{qrs}\,e^{q}\wedge
e^{r}\wedge e^{s}$ can be written as $\Omega _{qrs}=w^{p}\varepsilon _{pqrs}$%
, and its dual vector is 
\begin{equation}
w^{t}=-\frac{1}{3!}\,\varepsilon ^{tqrs}\,\Omega _{qrs},
\label{eq:hodgegeneral}
\end{equation}%
where $\varepsilon ^{tqrs}=\eta ^{tt^{\prime }}\eta ^{qq^{\prime }}\eta
^{rr^{\prime }}\eta ^{ss^{\prime }}\varepsilon _{t^{\prime }q^{\prime
}r^{\prime }s^{\prime }}$ is obtained from $\varepsilon _{0123}=+1$ with the
Lorentzian metric, so that $\varepsilon ^{tqrs}\varepsilon
_{pqrs}=-3!\,\delta _{p}^{t}$. All quantities below are therefore ordinary
tensors; projecting the Lorentz indices with the vierbein introduces no
factors of $\sqrt{-g}$. Applying Eq.~(\ref{eq:hodgegeneral}) term by term:

\begin{enumerate}
\item[(i)] \textbf{The Dynamic Sector.} With $R^{bc}=\frac{1}{2}%
R^{bc}{}_{qr}\,e^{q}\wedge e^{r}$, the dual of $\frac{1}{2}\varepsilon
_{abcd}R^{bc}\wedge e^{d}$ is $\frac{1}{4}\delta _{abc}^{tqr}R^{bc}{}_{qr}$,
where $\delta _{abc}^{tqr}$ is the generalized Kronecker delta. Expanding
it, 
\begin{equation}
\Omega _{a}=\frac{1}{2}\varepsilon _{abcd}R^{bc}\wedge
e^{d}\;\longrightarrow \;w^{t}=\frac{1}{2}\delta
_{a}^{t}R-R^{t}{}_{a}=-G_{(T)a}{}^{t},  \label{10a}
\end{equation}%
with $R^{t}{}_{a}=R^{tc}{}_{ac}$ and $R=R^{a}{}_{a}$. This is the standard
Einstein $3$-form of first-order gravity; its source is the ordinary
energy-momentum tensor. In the presence of torsion $G_{(T)ab}$ need not be
symmetric.

\item[(ii)] \textbf{The Kinematic Sector.} For the volume term, $\Omega
_{qrs}=-(3/l^{2})\varepsilon _{aqrs}$, and Eq.~(\ref{eq:hodgegeneral}) gives 
\begin{equation}
\Omega _{a}=-\frac{1}{2l^{2}(x)}\varepsilon _{abcd}e^{b}\wedge e^{c}\wedge
e^{d}\;\longrightarrow \;w^{t}=-\frac{3}{l^{2}(x)}\delta _{a}^{t}=-\Lambda
(x)\,\delta _{a}^{t},  \label{10b}
\end{equation}%
with $\Lambda (x)=3/l^{2}(x)$. This term characterizes the intrinsic
background geometry of the local tangent space rather than a propagating
gravitational force.

\item[(iii)] \textbf{The Nieh-Yan Sector.} Writing $T_{a}=\frac{1}{2}%
T_{ars}\,e^{r}\wedge e^{s}$, the components of the $3$-form are 
\begin{equation}
\left( 2\alpha \,dl(x)\wedge T_{a}\right) _{qrs}=6\alpha \,\partial
_{\lbrack q}l\,T_{|a|\,rs]},  \label{eq:resultado}
\end{equation}%
and its dual is 
\begin{equation}
\Omega _{a}=2\alpha \,dl(x)\wedge T_{a}\;\longrightarrow \;w^{t}=-\alpha
\,\varepsilon ^{tqrs}\,\partial _{q}l\,T_{ars}\equiv -\mathrm{NY}_{a}{}^{t}.
\label{eq:NYclosed}
\end{equation}

\item[(iv)] \textbf{The Matter Sector.} The sign of the correspondence
between ${}^{\ast }\tau _{a}$ and $\tau _{a}{}^{t}$ follows from the
definition~(\ref{9}). For a vacuum energy, $\mathcal{L}_{m}=-\frac{\rho _{v}%
}{4!}\varepsilon _{abcd}e^{a}\wedge e^{b}\wedge e^{c}\wedge e^{d}$, the
variation gives ${}^{\ast }\tau _{a}=\frac{\rho _{v}}{3!}\varepsilon
_{abcd}e^{b}\wedge e^{c}\wedge e^{d}$, whose dual is $+\rho _{v}\delta
_{a}^{t}$, while the corresponding energy-momentum tensor is $\tau
_{a}{}^{t}=-\rho _{v}\delta _{a}^{t}$. Hence 
\begin{equation}
\Omega _{a}=8\pi G\left( {}^{\ast }\tau _{a}-\frac{1}{4l^{2}(x)}{}^{\ast
}K_{a}\right) \;\longrightarrow \;w^{t}=-8\pi G\left( \tau _{a}{}^{t}-\frac{1%
}{4l^{2}(x)}K_{a}{}^{t}\right) .  \label{10c}
\end{equation}%
As a check, the same vacuum energy shifts the cosmological term by $8\pi
G\rho _{v}$, consistently with $\Lambda =3/l^{2}$ and $\kappa =1/16\pi G$.
\end{enumerate}

Collecting Eqs.~(\ref{10a})-(\ref{10c}), multiplying by $-1$ and projecting
with the vierbein, the de Sitter-Palatini field equation~(\ref{ecmov}) takes
the tensorial form 
\begin{equation}
G_{(T)}^{\mu \nu }+\Lambda (x)g^{\mu \nu }+\mathrm{NY}^{\mu \nu }=8\pi
G\left( T^{\mu \nu }-\frac{1}{4l^{2}(x)}K^{\mu \nu }\right) ,  \label{10d}
\end{equation}%
with 
\begin{equation}
\mathrm{NY}^{\mu \nu }=\alpha \,\varepsilon ^{\nu \lambda \rho \sigma
}\,\partial _{\lambda }l\;T^{\mu }{}_{\rho \sigma }.  \label{eq:NYmunu}
\end{equation}%
In coordinate components, $\varepsilon ^{\nu \lambda \rho \sigma }$ is the
Levi-Civita tensor, i.e.\ $1/\sqrt{-g}$ times the numerical symbol, up to
the sign fixed by $\varepsilon _{0123}=+1$; no other factor of $\sqrt{-g}$
appears, since all terms in Eq.~(\ref{10d}) are obtained with the same rule~(%
\ref{eq:hodgegeneral}).

Recalling the identity for the background de Sitter tensor, $G_{(K)}^{\mu
\nu }=-\Lambda (x)g^{\mu \nu }$ [Appendix~\ref{app:diffeo}, Eq.~(\ref{eq:B.9}%
)], Eq.~(\ref{10d}) reproduces the splitting $G_{(T)}^{\mu \nu
}-G_{(K)}^{\mu \nu }$ of Refs.~\cite{60}-\cite{73}, extended by the Nieh-Yan
term. This confirms that the subtraction of the dynamic and kinematic
Einstein tensors is not an ad hoc arrangement, but the tensor manifestation
of the gauge algebra in the local tangent space.

\section{Lorentz Variation and Torsion Dynamics}

\label{sec:5}

To turn on torsion explicitly and find the form of the matter spin tensor under the de Sitter group $SO(4,1)$, we vary the action with respect to the Lorentz connection $\omega ^{ab}$. This variation must be carried out without assuming any a priori mathematical restrictions.

Throughout this section every variation is written in the form $X_{ab}\wedge
\delta \omega ^{ab}$, with the $1$-form $\delta \omega ^{ab}$ placed to the
right, the same ordering used for $\delta e^{a}$ in Sec.~\ref{sec:4}.

(i) \textbf{Variation of the Gravitational Sector.} Since the kinematic term
of $S_{g}$ contains only vierbeins and the function $l(x)$, it does not
contribute to the variation with respect to the Lorentz connection. Applying
Palatini's identity $\delta R^{ab}=D(\delta\omega^{ab})$ to the first
integral of Eq.~(\ref{6}), integrating by parts, and using $\varepsilon
_{abcd}D(e^{c}\wedge e^{d})=2\varepsilon _{abcd}T^{c}\wedge e^{d}$, we find 
\begin{equation}
\delta_{\omega}S_{MEHC}=\kappa\int\varepsilon_{abcd}\,\delta\omega^{ab}%
\wedge T^{c}\wedge e^{d}=-\kappa\int\varepsilon_{abcd}T^{c}\wedge
e^{d}\wedge\delta\omega^{ab},  \label{eq:5.1}
\end{equation}
where the minus sign arises from moving the $1$-form $\delta \omega ^{ab}$
past the $1$-form $e^{d}$.

(ii) \textbf{Variation of the Nieh-Yan Term.} The Nieh-Yan coupling also
depends on $\omega^{ab}$ through $T^{a}=de^{a}+\omega^{a}{}_{b}\wedge e^{b}$%
, so that $\delta T^{a}=\delta \omega ^{a}{}_{b}\wedge e^{b}$, and therefore
the variation of the Nieh-Yan term [Eq.~(\ref{eq:SNY})] leads to 
\begin{equation}
\delta_{\omega}S_{NY}=2\alpha\kappa\int dl(x)\wedge e_{a}\wedge e_{b}\wedge
\delta\omega^{ab}.  \label{eq:domegaSNY}
\end{equation}

(iii) \textbf{Variation of the Matter Sector.} As before, and in the same $%
\hbar=c=1$ units as Eq.~(\ref{9}), 
\begin{equation}
\delta_{\omega}S_{m}=\int{}^{\ast}S_{ab}\wedge\delta\omega ^{ab}.
\label{eq:5.2}
\end{equation}

\begin{enumerate}
\item[(iv)]  \textbf{The de Sitter-Cartan Equation for Torsion.}
Requiring the principle of total stationarity for the Lorentz connection $%
\delta _{\omega }(S_{g}+S_{NY}+S_{m})=0$ under any arbitrary and independent
variation $\delta \omega ^{ab}$, we arrive at the following equation, 
\begin{equation}
\frac{1}{2}\varepsilon _{abcd}T^{c}\wedge e^{d}=8\pi G\,{}^{\ast
}S_{ab}+\alpha \,dl(x)\wedge e_{a}\wedge e_{b}.  \label{eq:5.4}
\end{equation}%
Here all three coefficients are antisymmetric in $a,b$, and we have divided by $2\kappa $, with $1/(2\kappa )=8\pi G$.

Torsion is now sourced both by matter spin and by the gradient of $l(x)$. To
express this equation in components, we write $T^{a}=\frac{1}{2}%
T^{a}{}_{bc}\,e^{b}\wedge e^{c}$ and ${}^{\ast }S_{ab}=-\frac{1}{3!}%
S^{p}{}_{ab}\,\varepsilon _{pqrs}\,e^{q}\wedge e^{r}\wedge e^{s}$, the sign
being fixed, as for ${}^{\ast }\tau _{a}$ in Eq.~(\ref{eq:taudual}), by the
definition~(\ref{7b}), and take the dual of both sides of Eq.~(\ref{eq:5.4}%
). The left-hand side yields the \emph{modified torsion} of Einstein-Cartan
theory, independently of the sign convention adopted for $\varepsilon
^{abcd} $, 
\begin{equation}
T^{\lambda }{}_{\mu \nu }+\delta _{\mu }^{\lambda }T^{\rho }{}_{\nu \rho
}-\delta _{\nu }^{\lambda }T^{\rho }{}_{\mu \rho }=-16\pi G\,S^{\lambda
}{}_{\mu \nu }-2\alpha \,\varepsilon ^{\lambda }{}_{\mu \nu \rho }\,\partial
^{\rho }l,  \label{eq:5.4b}
\end{equation}%
where Latin indices have been projected onto the coordinate manifold with
the vierbein, and $\varepsilon ^{\lambda }{}_{\mu \nu \rho }=g^{\lambda
\kappa }\varepsilon _{\kappa \mu \nu \rho }$ is obtained from $\varepsilon
_{0123}=+1$ by raising indices with the metric. Contracting $\lambda $ with $%
\mu $ gives $T^{\rho }{}_{\nu \rho }=8\pi G\,S^{\rho }{}_{\nu \rho }$ (the
Levi-Civita term is traceless), and substituting back we find 
\begin{equation}
T^{\lambda }{}_{\mu \nu }=-16\pi G\left( S^{\lambda }{}_{\mu \nu }+\frac{1}{2%
}\delta _{\mu }^{\lambda }S^{\rho }{}_{\nu \rho }-\frac{1}{2}\delta _{\nu
}^{\lambda }S^{\rho }{}_{\mu \rho }\right) -2\alpha \,\varepsilon ^{\lambda
}{}_{\mu \nu \rho }\,\partial ^{\rho }l.  \label{eq:5.5}
\end{equation}%
The trace terms enter with a relative minus sign, so that the right-hand
side is antisymmetric in $(\mu ,\nu )$, as required by $T^{\lambda }{}_{\mu
\nu }=-T^{\lambda }{}_{\nu \mu }$; their structure $\delta _{\mu }^{\lambda
}V_{\nu }-\delta _{\nu }^{\lambda }V_{\mu }$ is the trace (vector)
irreducible component of the torsion. Equation~(\ref{eq:5.4b}%
) has the same structure as Cartan's equation of Ref.~\cite{5}, in which the
modified torsion is proportional to the spin tensor. The overall factor $%
-16\pi G$ reflects the normalization of $S^{\lambda }{}_{\mu \nu }$ fixed by
Eq.~(\ref{7b}), in which $\delta \omega ^{ab}$ is summed over all $(a,b)$,
and its sign by the orientation $e^{0123}=e\,d^{4}x$, for which ${}^{\ast
}S_{ab}\wedge \delta \omega ^{ab}=S^{\mu }{}_{ab}\,\delta \omega
^{ab}{}_{\mu }\,e\,d^{4}x$. For the FLRW ansatz of Sec.~\ref{sec:6.2}, the
last term of Eq.~(\ref{eq:5.5}) gives $T^{i}{}_{jk}\supset -2\alpha \,\dot{l}%
\,\varepsilon _{ijk}$, i.e.\ a contribution $+2\alpha \dot{l}$ to the axial
torsion $q(t)$, Eq.~(\ref{eq:qalgebraic}). The Nieh-Yan
contribution is built from the Levi-Civita tensor, not the Kronecker delta
suggested by surface analogy with the Peccei-Quinn axion literature \cite%
{83,84}. This reflects that the Nieh-Yan term is parity-odd and topological,
so its variation produces an axial rather than a trace-type tensor
structure. Torsion remains algebraically determined, as in Einstein-Cartan
gravity \cite{5,6,7,8,9}: it is sourced jointly by the material spin density
and by the gradient of the dynamical de Sitter scale. However, because the axial part of the torsion is now fixed by $\partial l$,
substituting it back into the equation for $l(x)$, Eq.~(\ref{eq:leom}),
turns that algebraic equation into a second-order equation for $l(x)$ (see Sec.~\ref{sec:6.2}). This is the same mechanism by which a field coupled to
the Nieh-Yan density acquires a kinetic term once torsion is integrated out 
\cite{83,84}. The geometric significance of the torsional structure, in its
original Riemann-Cartan formulation, is discussed at length in Refs.~\cite%
{88,89}.
\end{enumerate}

\section{De Sitter conservation law and the torsion-$\Lambda$ coupling}

\label{sec:6}

We now consider a test of the global gauge consistency of the system by
taking the exterior covariant derivative $\mathcal{D}$ of the complete de
Sitter-Palatini field equation (\ref{ecmov}). The curvature and torsion obey
the Lorentz Bianchi identities $\mathcal{D}R^{ab}=0$ and $\mathcal{D}%
T^{a}=R^{a}{}_{b}\wedge e^{b}$; for constant $l$ these are equivalent to the 
$\mathfrak{so}(4,1)$ identity $\mathcal{D}F=0$, and the terms containing $dl$
below measure the departure from it.

We apply the exterior covariant derivative $\mathcal{D}$ to both sides of Eq.~(\ref{ecmov}), keeping the free Lorentz index in the first position. Applying the Leibniz rule to the
kinematic volume term without suppressing the torsion field ($T^{c}=\mathcal{%
D}e^{c}$), one finds 
\begin{equation}
\mathcal{D}\left[ \frac{1}{2}\varepsilon _{abcd}R^{bc}\wedge e^{d}\right] -%
\mathcal{D}\left[ \frac{1}{2l^{2}(x)}\varepsilon _{abcd}e^{b}\wedge
e^{c}\wedge e^{d}\right] +\mathcal{D}\left[ 2\alpha \,dl(x)\wedge T_{a}%
\right] =8\pi G\,\mathcal{D}^{\ast }\Pi _{a},  \label{13a}
\end{equation}%
where, using $\mathcal{D}R^{bc}=0$, $\mathcal{D}T_{a}=R_{ab}\wedge e^{b}$
and $\varepsilon _{abcd}\mathcal{D}(e^{b}\wedge e^{c}\wedge
e^{d})=3\varepsilon _{abcd}e^{b}\wedge e^{c}\wedge T^{d}$, 
\begin{align}
\mathcal{D}\left[ \frac{1}{2}\varepsilon _{abcd}R^{bc}\wedge e^{d}\right] & =%
\frac{1}{2}\varepsilon _{abcd}R^{bc}\wedge T^{d},  \notag \\
\mathcal{D}\left[ \frac{1}{2l^{2}(x)}\varepsilon _{abcd}e^{b}\wedge
e^{c}\wedge e^{d}\right] & =d\left( \frac{1}{2l^{2}(x)}\right) \wedge
\varepsilon _{abcd}e^{b}\wedge e^{c}\wedge e^{d}+\frac{3}{2l^{2}(x)}%
\varepsilon _{abcd}e^{b}\wedge e^{c}\wedge T^{d},  \notag \\
\mathcal{D}\left[ 2\alpha \,dl(x)\wedge T_{a}\right] & =-2\alpha
\,dl(x)\wedge \mathcal{D}T_{a}=-2\alpha \,dl(x)\wedge R_{ab}\wedge e^{b}.
\label{13b}
\end{align}%
Regrouping the terms, one finds the de Sitter conservation law of $4$-forms: 
\begin{equation}
\begin{split}
&-d\left( \frac{1}{2l^{2}(x)}\right) \wedge \varepsilon _{abcd}e^{b}\wedge
e^{c}\wedge e^{d}+\varepsilon _{abcd}\left( \frac{1}{2}R^{bc}-\frac{3}{%
2l^{2}(x)}e^{b}\wedge e^{c}\right) \wedge T^{d} \\
&\qquad -2\alpha \,dl(x)\wedge R_{ab}\wedge e^{b}=8\pi G\,\mathcal{D}^{\ast
}\Pi _{a}.
\end{split}
\label{13c}
\end{equation}

The new term, $-2\alpha\,dl(x)\wedge R_{ab}\wedge e^{b}$, couples $dl(x)$
directly to the Riemann curvature, not just to the torsion. It is not only
"torsion sourced by $dl$"; the consistency identity itself now links $dl(x)$
to the background curvature (in FLRW, through the torsion-induced part $%
(H-h)q$ of $R^{0i}$, Sec.~\ref{sec:6.2}). This is consistent with $S_{NY}$,
via Nieh-Yan, reducing to $T^{a}T_{a}-e^{a}e^{b}R_{ab}$, where the curvature
term was present from the beginning, only "hidden" in the topological part
before taking $\mathcal{D}$.

Equation (\ref{13c}) is the integrability condition of the
field equations (\ref{ecmov}). Its geometric terms do not vanish
identically, but only once the torsion equation and Eq.~(\ref{eq:leom}) are
used, leaving the conservation law of the source, just as the ordinary Bianchi identity $\nabla _{\mu }G^{\mu \nu }=0$ implies $\nabla
_{\mu }T^{\mu \nu }=0$ in standard cosmology. Since ${}^{\ast }\Pi _{a}$
contains $-{}^{\ast }K_{a}/4l^{2}$, its covariant derivative includes the
term $-d(1/4l^{2})\wedge {}^{\ast }K_{a}$, which couples $dl$ to the
conformal current. It will be used as a consistency check on the resulting
solution.

\subsection{Spacetime with Variable $\Lambda(x)$}

As established in Sec.~\ref{sec:4.1}, a torsion sector of purely trace type,
sourced by the trace part $\Sigma (t)$ of the spin density, gives $\mathcal{N%
}(x)\equiv 0$ and cannot activate the dynamics of $l(x)$; an axial
(parity-odd) torsion component is required. As shown below, such a component
is generated by the variation of $l(x)$ itself, through the Nieh-Yan term in
the torsion equation, and may in addition be sourced by a chiral spin
density. We therefore consider a torsion ansatz including both channels,
Eq.~(\ref{eq:6.6}) below.

\subsection{An Explicit, Verified Cosmological Solution}

\label{sec:6.2}

We adopt the spatially flat FLRW coframe $e^{0}=dt$, $e^{i}=a(t)dx^{i}$ ($%
i=1,2,3$), together with torsion of both trace and axial type, 
\begin{equation}
T^{0}=0,\qquad T^{i}=h(t)\,e^{0}\wedge e^{i}-\frac{1}{2}q(t)\,\varepsilon
^{i}{}_{jk}\,e^{j}\wedge e^{k},  \label{eq:6.6}
\end{equation}
where $h(t)$ is sourced by the trace spin density $\Sigma(t)$, and $q(t)$ is
sourced by $\dot{l}$ and, possibly, by a chiral spin density $\Sigma_{A}(t)$%
; it is this axial component that activates Eq.~(\ref{eq:leom}).

\textbf{Connection and curvature.} Using the ansatz~(\ref{eq:6.6}) in the
Cartan structure equation, we find 
\begin{equation}
\omega ^{0i}=(H-h)\,e^{i},\qquad \omega ^{ij}=\frac{1}{2}q\,\varepsilon
^{ijk}\,e_{k},  \label{eq:6.conn}
\end{equation}%
with $H\equiv \dot{a}/a$. The resulting curvature 2-forms are 
\begin{align}
R^{0i}& =\left[ \dot{H}-\dot{h}+H(H-h)\right] e^{0}\wedge e^{i}-\frac{1}{2}%
(H-h)q\,\varepsilon ^{i}{}_{jk}\,e^{j}\wedge e^{k}, \\
R^{ij}& =\left[ (H-h)^{2}-\frac{q^{2}}{4}\right] e^{i}\wedge e^{j}+\frac{1}{2%
}(\dot{q}+Hq)\,\varepsilon ^{ij}{}_{k}\,e^{0}\wedge e^{k}.
\end{align}%
The corresponding Ricci tensor, computed directly from this curvature,
reproduces the standard torsion-free FLRW result in the limit $h=q=0$: $%
R_{00}=-3(\dot{H}+H^{2})$ and $R_{ij}/a^{2}\delta _{ij}=\dot{H}+3H^{2}$,
matching an independent Christoffel-symbol computation and confirming the
internal consistency of the connection and curvature above.

\textbf{The Nieh-Yan invariant.} Direct computation gives 
\begin{equation}
T^{a}\wedge T_{a}=-6hq\,e^{0123},\qquad e^{a}\wedge e^{b}\wedge
R_{ab}=\left( -6hq+3\dot{q}+9Hq\right) e^{0123},
\end{equation}%
so that the trace-torsion contributions cancel, leaving 
\begin{equation}
\mathcal{N}(t)=-3\left( \dot{q}+3Hq\right) =-\frac{3}{a^{3}}\frac{d}{dt}%
\left( a^{3}q\right) .  \label{eq:Nresult}
\end{equation}%
This total comoving derivative confirms that the trace sector ($h$, trace
spin) is decoupled from Eq.~(\ref{eq:leom}): only the axial torsion sources $%
l(x)$.

\textbf{Field equations for the ansatz.} Substituting the ansatz into the
vierbein equation~(\ref{ecmov}), the torsion equation~(\ref{eq:5.4}) and the 
$l$-equation~(\ref{eq:leom}), the only non-vanishing components are the
following.

\textbf{Matter sector.} Inverting the dual map of Sec.~\ref{sec:4.2}, with
the direct correspondence~(\ref{10c}), the matter $3$-form is expressed in
terms of the energy-momentum tensor as 
\begin{equation}
{}^{\ast }\tau _{a}=-\frac{1}{3!}\,\tau _{a}{}^{p}\,\varepsilon
_{pqrs}\,e^{q}\wedge e^{r}\wedge e^{s}.  \label{eq:taudual}
\end{equation}%
For a vacuum energy, $\tau _{a}{}^{p}=-\rho _{v}\delta _{a}^{p}$, this gives 
${}^{\ast }\tau _{a}=\frac{\rho _{v}}{3!}\varepsilon _{abcd}e^{b}\wedge
e^{c}\wedge e^{d}$, which is the variation of $\mathcal{L}_{m}=-\frac{\rho
_{v}}{4!}\varepsilon _{abcd}e^{a}\wedge e^{b}\wedge e^{c}\wedge e^{d}$,
confirming the sign. For a perfect fluid comoving with the frame, $%
u^{a}=\delta _{0}^{a}$, one has $\tau ^{ab}=(\rho +p)u^{a}u^{b}+p\,\eta
^{ab} $, i.e. $\tau _{0}{}^{0}=-\rho $ and $\tau _{i}{}^{i}=p$ (no sum), all
other components vanishing. Eq.~(\ref{eq:taudual}) then gives 
\begin{equation}
{}^{\ast }\tau _{0}=\frac{\rho }{3!}\,\varepsilon _{0ijk}\,e^{i}\wedge
e^{j}\wedge e^{k}=\rho \;e^{1}\wedge e^{2}\wedge e^{3},\qquad {}^{\ast }\tau
_{1}=p\;e^{0}\wedge e^{2}\wedge e^{3},  \label{eq:taufluid}
\end{equation}%
and cyclically for $a=2,3$. On the left-hand side of Eq.~(\ref{ecmov}), the $%
a=0$ component is proportional to $e^{1}\wedge e^{2}\wedge e^{3}$ and the $%
a=i$ components to $e^{0}\wedge e^{j}\wedge e^{k}$, with $\varepsilon
_{ijk}=1$; equating coefficients of the same $3$-forms on both sides, the
vierbein equation has only diagonal components, 
\begin{align}
\mathbf{(}00\mathbf{):}\qquad & 3(H-h)^{2}-\frac{3}{4}q^{2}-\frac{3}{l^{2}}%
=8\pi G\rho ,  \label{eq:final00} \\
\mathbf{(}ii\mathbf{):}\qquad & -h^{2}+\frac{1}{4}q^{2}+2\dot{h}+\frac{3}{%
l^{2}}+4hH-2\frac{\ddot{a}}{a}-H^{2}-2\alpha q\dot{l}=8\pi Gp,
\label{eq:finalii}
\end{align}%
which reduce to the Friedmann and Raychaudhuri equations for $h=q=0$, $%
l\rightarrow \infty $. In writing them we have neglected the
conformal-current contribution $K^{\mu \nu }/4l^{2}$. Since $\vartheta _{\nu
}^{\mu }$ is built from the coordinates $x^{\mu }$, it is not compatible
with homogeneity in a global FLRW chart. Within a local patch around the
observer it is suppressed by $x^{2}/l^{2}$, and its covariant cosmological
treatment is left for future work. In vacuum, $\rho =p=0$, it vanishes
identically, so that the vacuum results below are exact. The Nieh-Yan term
contributes only to the $(ii)$ component, through the term $-2\alpha q\dot{l}
$. In the de Sitter limit $h=q=0$, $\dot{l}=0$, both Eqs.~(\ref{eq:final00})
and~(\ref{eq:finalii}) give $H^{2}=1/l^{2}=\Lambda /3$. The torsion
equation~(\ref{eq:5.4}) splits into a mixed and a purely spatial sector, 
\begin{align}
(a,b)=(0,i):\qquad & h(t)=\kappa _{h}\Sigma (t),  \label{eq:htrace} \\
(a,b)=(i,j):\qquad & q(t)=\kappa ^{\prime }\Sigma _{A}(t)+2\alpha \dot{l}(t),
\label{eq:qalgebraic}
\end{align}%
where $\Sigma $ and $\Sigma _{A}$ are the trace and axial components of the
spin density and $\kappa _{h}$, $\kappa ^{\prime }$ collect the constants of
Eq.~(\ref{eq:5.4}). The Nieh-Yan term contributes only to the axial channel.
Finally, with Eq.~(\ref{eq:Nresult}), Eq.~(\ref{eq:leom}) becomes 
\begin{equation}
\mathbf{(}l(x)\mathbf{):}\qquad \frac{1}{l^{3}(t)}=\frac{\alpha }{2}\left( 
\dot{q}+3Hq\right) .  \label{eq:finalleq}
\end{equation}

\textbf{Consistency.} Equations~(\ref{eq:final00})-(\ref{eq:finalleq}) are
not independent: Eq.~(\ref{eq:final00}) is a constraint, and its time
derivative must follow from the others. Computing $\frac{d}{dt}$(\ref%
{eq:final00}) and eliminating $\ddot{a}$ with Eq.~(\ref{eq:finalii}) and $%
\ddot{l}$ with Eqs.~(\ref{eq:qalgebraic})-(\ref{eq:finalleq}), one finds,
for $h=0$, 
\begin{equation}
\dot{\rho}+3H(\rho +p)=-\frac{3\kappa ^{\prime }}{16\pi G}\,\Sigma
_{A}\left( \dot{q}+Hq\right) .  \label{eq:exchange}
\end{equation}%
For $\Sigma _{A}=0$ the system is therefore consistent with
the standard conservation of the perfect fluid, $\dot{\rho}+3H(\rho +p)=0$.
A non-vanishing axial spin density exchanges energy with the fluid, and its
own evolution must then be supplied by a dynamical model of the spin fluid.
All the expressions above were obtained by direct computer-algebra
evaluation of the exterior-form equations on the ansatz~(\ref{eq:6.6}), and
the consistency relation~(\ref{eq:exchange}) was checked both symbolically
and along numerically integrated trajectories (residuals $\lesssim 10^{-14}$%
).

\textbf{The dynamics of $l(t)$.} In the absence of spin sources, $h=\Sigma
_{A}=0$, Eq.~(\ref{eq:qalgebraic}) gives $q=2\alpha \dot{l}$: the axial
torsion is generated by the time variation of the de Sitter scale.
Substituting into Eqs.~(\ref{eq:final00}), (\ref{eq:finalii}) and~(\ref%
{eq:finalleq}), the system closes on $a(t)$ and $l(t)$: 
\begin{align}
H^{2}& =\alpha ^{2}\dot{l}^{2}+\frac{1}{l^{2}}+\frac{8\pi G}{3}\rho ,
\label{eq:Fried} \\
\ddot{l}+3H\dot{l}& =\frac{1}{\alpha ^{2}\,l^{3}},  \label{eq:ldampedosc} \\
\dot{H}& =-3\alpha ^{2}\dot{l}^{2}-4\pi G(\rho +p).  \label{eq:Hdot}
\end{align}%
Thus Eq.~(\ref{eq:leom}), algebraic in $l$ as it stands, becomes a
second-order equation once the torsion equation is used: $l(x)$ acquires
genuine cosmological dynamics. Defining, in units $8\pi G=1$, 
\begin{equation}
\phi \equiv \sqrt{6}\,\alpha \,l,\qquad V(\phi )\equiv \frac{3}{l^{2}}=\frac{%
18\,\alpha ^{2}}{\phi ^{2}},  \label{eq:phiV}
\end{equation}%
Eqs.~(\ref{eq:Fried})-(\ref{eq:Hdot}) become $3H^{2}=\frac{1}{%
2}\dot{\phi}^{2}+V+\rho $, $\ddot{\phi}+3H\dot{\phi}+V^{\prime }(\phi )=0$
and $\dot{H}=-\frac{1}{2}\dot{\phi}^{2}-\frac{1}{2}(\rho +p)$. The local de
Sitter scale behaves as a canonical quintessence field with an inverse
power-law potential $V\propto \phi ^{-2}$, i.e.\ the Ratra-Peebles model 
\cite{90} with index $n=2$. The local cosmological parameter $\Lambda
(t)=3/l^{2}(t)$ is the potential energy $V(\phi )$.

This identification is not restricted to FLRW. The connection field equation
is algebraic, so the torsion can be eliminated from the action. In the
absence of spin sources the resulting torsion is axial and proportional to $%
\alpha \,\partial l$, Eq.~(\ref{eq:5.5}); substituted back, both the
contorsion-squared terms of the Einstein-Cartan action and the Nieh-Yan term
itself become quadratic in $\partial l$. Eliminating the torsion explicitly
(Appendix~\ref{app:Seff}) gives the covariant effective action 
\begin{equation}
S_{\mathrm{eff}}=\int \sqrt{-g}\,\Big[\frac{\mathring{R}}{16\pi G}-\frac{1}{2%
}Z(l)\,g^{\mu \nu }\partial _{\mu }l\,\partial _{\nu }l-V(l)\Big]+S_{m},
\label{eq:Seff}
\end{equation}%
where $\mathring{R}$ is the Riemannian scalar curvature, $Z=6\alpha
^{2}/8\pi G$ and $V=3/(8\pi G\,l^{2})$. These coefficients coincide with
those obtained above from the FLRW reduction, which is thus recovered from a
covariant action. The kinetic term has the healthy sign. The theory
therefore propagates the two degrees of freedom of General Relativity plus
one scalar, $l(x)$, in agreement with the fact that Eq.~(\ref{eq:ldampedosc}%
) requires both $l$ and $\dot{l}$ as initial data. This is the same
mechanism by which a scalar coupled to the Nieh-Yan density acquires a
kinetic term once torsion is integrated out \cite{83,84}. The limit $\alpha
\rightarrow 0$ is singular: the kinetic term vanishes and Eq.~(\ref{eq:leom}%
) degenerates into Eq.~(\ref{eq:degenerate}), consistent with the absence of
dynamics for $l$ without the Nieh-Yan coupling.

Two properties follow directly. First, $l(t)$ can never reach zero: the
quantity $E\equiv \alpha ^{2}\dot{l}^{2}+1/l^{2}$ obeys, by Eq.~(\ref%
{eq:ldampedosc}), 
\begin{equation}
\dot{E}=-6\alpha ^{2}H\dot{l}^{2}\leq 0,
\end{equation}%
for an expanding universe ($H>0$), so that $1/l^{2}\leq E(t_{i})$ and $%
l(t)\geq E(t_{i})^{-1/2}>0$ for all later times; the potential barrier $%
1/l^{2}$ reflects any initially decreasing $l(t)$. Second, at late times the
solution approaches the slow-roll attractor $3H\dot{l}\simeq 1/(\alpha
^{2}l^{3})$ with $H\simeq 1/l$, i.e. 
\begin{equation}
l^{3}(t)\simeq \frac{t}{\alpha ^{2}},\qquad \Lambda (t)\simeq 3\left( \frac{%
\alpha ^{2}}{t}\right) ^{2/3},\qquad w_{\mathrm{eff}}\equiv -1-\frac{2\dot{H}%
}{3H^{2}}\simeq -1+\frac{2}{9\alpha ^{2}l^{2}}\longrightarrow -1,
\label{eq:attractor}
\end{equation}%
so that $\Lambda (t)$ relaxes to zero while the expansion keeps accelerating.

\begin{figure}[t]
\centering
\includegraphics[width=0.95\textwidth]{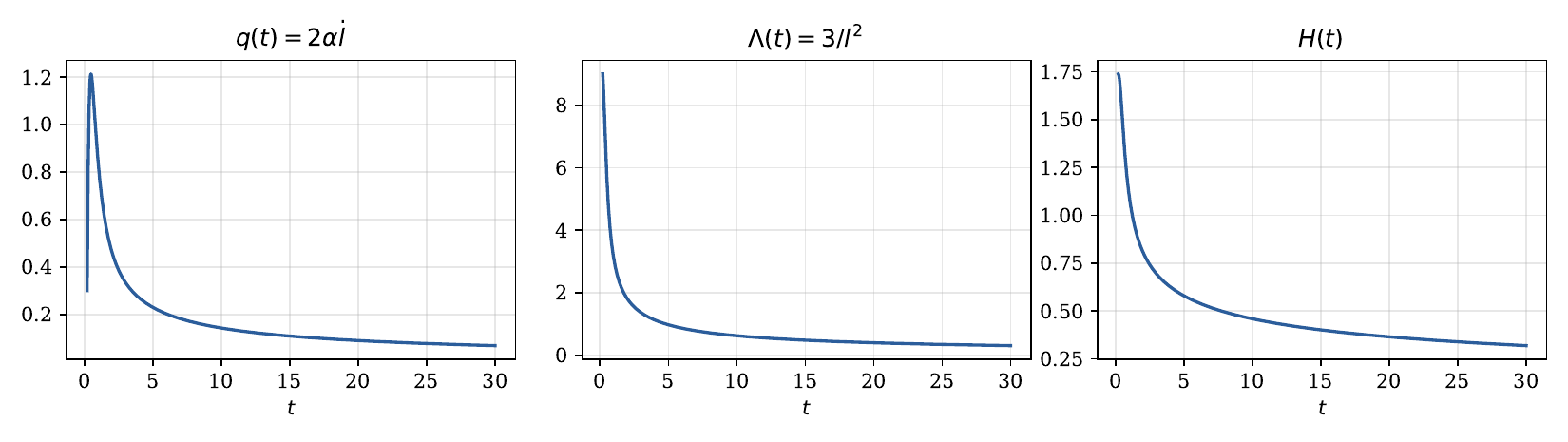}
\caption{The vacuum axial-torsion solution of Eqs.~(\protect\ref{eq:Fried})-(%
\protect\ref{eq:ldampedosc}), with $\protect\alpha=1$, $q(t_{i})=0.3$, $%
\Lambda(t_{i})=9$. Left: the axial torsion $q(t)=2\protect\alpha\dot{l}$.
Center: the local cosmological parameter $\Lambda (t)=3/l^{2}(t)$, relaxing
monotonically. Right: the expansion rate $H(t)$. The solution satisfies the
complete set of field equations, Eqs.~(\protect\ref{eq:final00})-(\protect
\ref{eq:finalleq}), including the torsion equation.}
\label{fig:axial}
\end{figure}

\textbf{The vacuum axial-torsion solution.} Integrating Eqs.~(\ref{eq:Fried}%
)-(\ref{eq:ldampedosc}) with $\rho =0$, $\alpha =1$ (arbitrary units), from $%
t_{i}=0.2$ to $t_{f}=30$, with $l(t_{i})=1/\sqrt{3}$ [$\Lambda (t_{i})=9$]
and $q(t_{i})=2\alpha \dot{l}(t_{i})=0.3$, gives the trajectory shown in
Fig.~\ref{fig:axial}. The axial torsion first grows, reaching $q\simeq 1.21$
at $t\simeq 0.48$, and then decays ($q=0.069$ at $t_{f}$); $\Lambda (t)$
decreases monotonically from $9$ to $0.30$, and $H(t)$ from $1.74$ to $0.32$%
. Extending the integration to $t=10^{6}$ confirms the attractor~(\ref%
{eq:attractor}), with $\alpha ^{2}l^{3}/t=1.00005$ and $w_{\mathrm{eff}%
}=-1.0000$. Along the whole trajectory the $(ii)$ equation and the $l$%
-equation are satisfied to machine precision, the $(00)$ constraint being
imposed.

\subsection{Matter and Late-Time Behaviour}

\label{sec:matter}

\begin{figure}[t]
\centering
\includegraphics[width=0.95\textwidth]{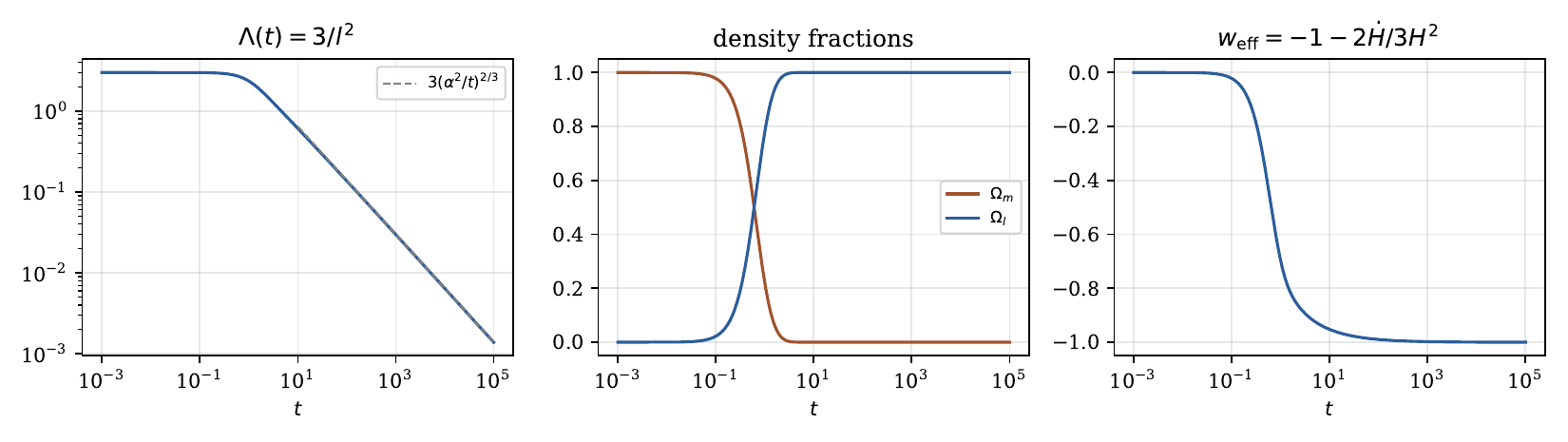}
\caption{Matter plus dynamical de Sitter scale, Eqs.~(\protect\ref{eq:Fried}%
)-(\protect\ref{eq:Hdot}) with $p=0$, $\protect\alpha=1$, $8\protect\pi G=1$%
, starting deep in matter domination with $\dot{l}=0$. Left: $\Lambda(t)$,
frozen during matter domination and then approaching the attractor $3(%
\protect\alpha^{2}/t)^{2/3}$ (dashed). Center: matter fraction $\Omega_{m}=%
\protect\rho/3H^{2}$ and de Sitter-scale fraction $\Omega_{l}=1-\Omega_{m}$.
Right: effective equation of state.}
\label{fig:dust}
\end{figure}

Including pressureless matter, $p=0$, $\rho \propto a^{-3}$, Eqs.~(\ref%
{eq:Fried})-(\ref{eq:Hdot}) describe the standard thawing-quintessence
history (Fig.~\ref{fig:dust}). During the matter-dominated phase, Hubble
friction in Eq.~(\ref{eq:ldampedosc}) keeps $l(t)$ essentially
frozen and $\Lambda (t)$ nearly constant. Once $\rho $ has diluted below $%
3/l^{2}$, the de Sitter scale starts to roll, $\Lambda (t)$ decreases as in
Eq.~(\ref{eq:attractor}), the matter fraction tends to zero and $w_{\mathrm{%
eff}}\rightarrow -1$. Since the force in Eq.~(\ref{eq:ldampedosc}) is
repulsive, no finite-time collapse $l\rightarrow 0$ occurs, and no
persistent source is required for late-time stability.

A chiral spin density $\Sigma _{A}(t)$, i.e.\ an imbalance
between left- and right-handed fermion number densities, is therefore not
needed to activate or stabilize the dynamics of $l(x)$. The axial torsion
required by Eq.~(\ref{eq:leom}) is generated by $\dot{l}$ itself through
Eq.~(\ref{eq:qalgebraic}). If present, $\Sigma _{A}$ acts as an additional
source of axial torsion and exchanges energy with ordinary matter according
to Eq.~(\ref{eq:exchange}). Standard Model chirality-flipping interactions become faster than the Hubble rate below $\sim 80\ \mathrm{TeV}$ \cite{91}. Such a contribution would therefore be relevant only
in the early Universe, where primordial chiral asymmetries can be sourced by
CP-violating processes, sphaleron transitions, or the chiral magnetic effect 
\cite{91}. It plays no role in the late-time behaviour described above.

\section{Conclusions}

The reformulation of de Sitter relativity within a first-order gauge
framework, developed throughout this work, offers a possible approach to the
tension between local kinematical symmetries and macroscopic gravitational
dynamics. The theory replaces the geometric constraints of the classical metric approach with a connection 1-form defined directly on the $\mathfrak{so}(4,1)$ Lie algebra. Through a modified Einstein-Hilbert-Cartan action supplemented by a topological Nieh-Yan coupling, it incorporates a localized, variable cosmological parameter $\Lambda (x)=3/l^{2}(x)$ derived from the gauge structure. This parameter is governed by an independent field equation.

\textbf{Relation to Refs.~\cite{60}-\cite{73}, \cite{75}:}
The dynamical equivalence between a variable de Sitter scale and General
Relativity, established for a MacDowell-Mansouri-type action by Ref.~\cite%
{73} as a Weyl-gauge identity, applies directly to our construction, because our action (\ref{6}) is of the same type. This is in contrast to the teleparallel construction of Ref.~\cite{72}, in which $\Lambda (x)$ is instead endowed
with independent dynamics through a different mechanism. The broader
polynomial family of $SO(1,4)$ actions studied by Ref.~\cite{75} shows that
both behaviours are possible in principle. The contributions of the present work are:

(i) an explicit first-order, Cartan-geometric derivation of the torsion
sector consistent with that equivalence, expressed in the
vierbein/spin-connection language natural for an eventual coupling to
fermionic matter.

(ii) a topological Nieh-Yan coupling that supplies $l(x)$ with the
independent equation of motion it previously lacked.

(iii) the complete reduction of the resulting spin-torsion-$\Lambda $ system
on an FLRW background (Sec.~\ref{sec:6.2}), in which the de Sitter scale
acquires second-order dynamics equivalent to Ratra-Peebles quintessence.

\textbf{Variational Consistency:} We have shown that the
combination of the dynamic and kinematic Einstein sectors, $%
G_{(T)}^{\mu\nu}-G_{(K)}^{\mu\nu},$ with the Nieh-Yan current in the coordinate manifold is not an arbitrary construction. It corresponds to the tensorial form of the local de Sitter gauge curvature, Eq.~(\ref{10d}).

\textbf{The In\"{o}n\"{u}-Wigner Contraction and the Large-$l$
Limit:} The transition from the de Sitter gauge connection to the flat
kinematics of Special Relativity remains governed by the In\"{o}n\"{u}%
-Wigner contraction of the Lie algebra $\mathfrak{so}(4,1)$ (see appendix~%
\ref{app:ads}). This structure is unaffected by the Nieh-Yan extension.

\textbf{The Conformal and Axial Origins of $\Lambda (x)$:}
The generalized Noether theorem on the local tangent space indicates that
the covariantly conserved source under de Sitter relativity is the current $\Pi ^{\mu \nu }$. This current couples the standard energy-momentum tensor to the proper conformal current of matter $K^{\mu \nu }$, and it sources the curvature $G_{(T)}^{\mu \nu }$ once $l(x)$ is already fixed,
through the vierbein equation~(\ref{10d}). The value of $l(x)$ itself, and
hence $\Lambda (x)=3/l^{2}(x)$, is governed by a separate equation, Eq.~(\ref%
{eq:leom}), sourced solely by the axial (Nieh-Yan) invariant of the torsion, 
$\mathcal{N}(x)$, Eq.~(\ref{eq:NYcomp}). Since the torsion equation fixes
the axial torsion in terms of $\partial l$, this equation is effectively a
second-order evolution equation for $l(x)$.

\textbf{De Sitter General Relativity:} The transition to de Sitter
relativity modifies the Strong Equivalence Principle, so that the local
vacuum responds to the local matter distribution rather than remaining
fixed. Through the Nieh-Yan coupling, the de Sitter scale $l(x)$ itself
becomes dynamical: on an FLRW background, eliminating the torsion turns it
into a canonical scalar field with potential $V=\Lambda=3/l^{2}$, i.e.\ an
inverse-square Ratra-Peebles potential \cite{90}.

\textbf{Cosmological Behaviour:} The reduced FLRW system is consistent with
the standard conservation of ordinary matter. The de Sitter scale is bounded
away from zero, remains frozen by Hubble friction during matter domination,
and at late times approaches the attractor $l^{3}\simeq t/\alpha ^{2}$, so
that $\Lambda (t)\propto t^{-2/3}$ relaxes to zero while the expansion
accelerates with $w_{\mathrm{eff}}\rightarrow -1$. No chiral spin density is
required: the axial torsion that activates the dynamics is generated by $%
\dot{l}$ itself. A primordial chiral asymmetry, if present, would act as an
additional source exchanging energy with matter, relevant only above the
Standard-Model chirality-flipping scale $T\sim 80$ TeV \cite{91}.
Confronting the thawing behaviour described here with observational
constraints on quintessence, and extending the analysis to perturbations,
are left for future work. Static, spherically symmetric configurations with
a dynamical $l(x)$ are a natural next step: since $l$ adds a propagating
scalar, Birkhoff's theorem is not expected to hold in general, in contrast
with the constant-$l$ and pre-geometric cases \cite{62,92}.

\acknowledgments

This work was supported in part by ANID-Chile through Fondecyt grant No.
1262414 and by Vicerrector\'{\i}a de Investigaci\'{o}n e Innovaci\'{o}n
through grant UNAP VRII No. 091/25. The author would like to thank S. Caro,
J. D\'{\i}az, D. Molina, V.C. Orozco and C. Vera for useful suggestions and
comments.

\noindent\textit{AI disclosure statement.} The author acknowledges the use of generative AI tools (Anthropic, Claude
Opus models accessed via Claude Code, 2026) in the preparation of this work. These tools were used to cross-check analytical derivations and a numerical example, to verify bibliographic references, to improve the clarity of the manuscript's wording, and to assist with LaTeX formatting.

All AI-assisted derivations were independently verified by the author. All
AI-assisted text, equations, and references were reviewed and, where
necessary, revised by the author, who takes full responsibility for the
content of this article.

\noindent\textbf{Data Availability Statement} This manuscript has no associated
data. [Author's comment: This is a theoretical study and no experimental data
have been generated or analysed.]

\smallskip
\noindent\textbf{Code Availability Statement} This manuscript has no associated
code/software. [Author's comment: The symbolic and numerical checks mentioned
in the text can be provided by the author upon reasonable request.]

\appendix

\section{The de Sitter group}

\label{app:ads}

The de Sitter group in four dimensions is the group of linear
transformations of a five-dimensional ambient space that leave invariant the
quadratic form 
\begin{equation}
\eta
_{AB}X^{A}X^{B}=-X_{0}^{2}+X_{1}^{2}+X_{2}^{2}+X_{3}^{2}+X_{4}^{2}=l^{2},
\label{ds1}
\end{equation}%
with $A,B=0,1,2,3,4$. The ambient metric is 
\begin{equation}
\eta _{AB}=%
\begin{pmatrix}
\eta _{ab} & 0 \\ 
0 & +1%
\end{pmatrix}%
,\qquad \eta _{ab}=\mathrm{diag}(-1,1,1,1),  \label{ds3}
\end{equation}%
with $a,b=0,1,2,3$, so that the group is $SO(4,1)$. (The anti-de Sitter
group $SO(3,2)$ is obtained by replacing $+1$ with $-1$ in the last entry.)

The group has ten generators $J_{AB}=-J_{BA}$, which in the vector
representation read $(J_{AB})^{C}{}_{D}=\delta _{A}^{C}\eta _{BD}-\delta
_{B}^{C}\eta _{AD}$. They decompose as $J_{AB}=\{L_{ab},\pi _{a}\}$, with $%
L_{ab}=J_{ab}$ and $\pi _{a}=J_{a4}=-J_{4a}$, and satisfy 
\begin{equation}
\left[ J_{AB},J_{CD}\right] =\eta _{AD}J_{BC}-\eta _{BD}J_{AC}-\eta
_{AC}J_{BD}+\eta _{BC}J_{AD},  \label{ds7}
\end{equation}%
from which 
\begin{eqnarray}
\left[ L_{ab},L_{cd}\right] &=&\eta _{ad}L_{bc}-\eta _{bd}L_{ac}-\eta
_{ac}L_{bd}+\eta _{bc}L_{ad},  \label{ds9} \\
\left[ L_{ab},\pi _{c}\right] &=&\eta _{bc}\pi _{a}-\eta _{ac}\pi _{b},
\label{ds10} \\
\left[ \pi _{a},\pi _{b}\right] &=&-\eta _{44}L_{ab}=-L_{ab},  \label{ds11}
\end{eqnarray}%
in agreement with Eq.~(\ref{p6}). This algebra is semisimple. The Poincar%
\'{e} algebra, which is not, is obtained by the In\"{o}n\"{u}-Wigner
contraction: setting $\pi _{a}=lP_{a}$ and taking $l\rightarrow \infty $ ($%
\Lambda \rightarrow 0$), Eq.~(\ref{ds11}) becomes $[P_{a},P_{b}]=0$.

The $\mathfrak{so}(4,1)$-valued connection is $A=\frac{1}{2}W^{AB}J_{AB}$,
with 
\begin{equation*}
W^{AB}=%
\begin{pmatrix}
\omega ^{ab} & e^{a}/l \\ 
-e^{b}/l & 0%
\end{pmatrix}%
,
\end{equation*}%
i.e.\ $W^{ab}=\omega ^{ab}$ and $W^{a4}=-W^{4a}=e^{a}/l$. Since $%
W^{a4}J_{a4}=W^{4a}J_{4a}=(e^{a}/l)\pi _{a}$, 
\begin{equation}
A=\frac{1}{2}\omega ^{ab}L_{ab}+\frac{e^{a}}{l}\pi _{a}.  \label{ds12}
\end{equation}%
In the vector representation $A^{C}{}_{D}=W^{C}{}_{D}$, and the curvature $%
F^{C}{}_{D}=dW^{C}{}_{D}+W^{C}{}_{E}\wedge W^{E}{}_{D}$ has components 
\begin{equation}
F^{ab}=R^{ab}+W^{a}{}_{4}\wedge W^{4b}=R^{ab}-\frac{\eta _{44}}{l^{2}}%
e^{a}\wedge e^{b},\qquad F^{a4}=\frac{1}{l}\left( de^{a}+\omega
^{a}{}_{b}\wedge e^{b}\right) =\frac{1}{l}T^{a},  \label{ds13}
\end{equation}%
which, for $\eta _{44}=+1$, reproduces Eq.~(\ref{3a}). The de Sitter
spacetime, $R^{ab}=e^{a}\wedge e^{b}/l^{2}$ with $T^{a}=0$, is the vacuum $%
F=0$ of this connection.

\section{Diffeomorphism Invariance and the Splitting of the Einstein Tensor}

\label{app:diffeo}

To derive the unified current from the coordinate manifold, we consider a
local spacetime displacement defined by a vector field $\zeta^{\mu}(x)=\xi_{%
\rho}^{\mu}\varepsilon^{\rho}(x)+\lambda^{\mu\nu}(x)x_{\nu}$, where $%
\xi_{\rho}^{\mu}=\delta_{\rho}^{\mu}-\vartheta_{\rho}^{\mu}/4l^{2}$ are the
Killing vectors of the de Sitter ``translation,'' with $\vartheta_{\rho
}^{\mu}=2\eta_{\rho\nu}x^{\nu}x^{\mu}-x^{2}\delta_{\rho}^{\mu}$ and $%
x^{2}=\eta_{\mu\nu}x^{\mu}x^{\nu}$. Under a general coordinate
transformation $x^{\mu}\longrightarrow x^{\prime\mu}=x^{\mu}+\zeta^{\mu}(x)$%
, the variation of the action $S=S_{g}+S_{m}$ (the Nieh-Yan term does not
contribute to this variation, being independent of $g_{\mu\nu}$ directly)
vanishes identically by diffeomorphism invariance ($\delta_{\xi}S=0$).

The variation of the matter action under a de Sitter ``translation'' is: 
\begin{equation}
\delta S_{m}=-\frac{1}{2}\int_{\Omega}\sqrt{-g}T_{\mu\nu}\delta
g^{\mu\nu}d^{4}x,\qquad\text{where}\qquad T_{\mu\nu}=-\frac{2}{\sqrt{-g}}%
\frac {\delta\mathcal{L}_{m}}{\delta g^{\mu\nu}},  \label{eq:B.1}
\end{equation}
is the energy-momentum tensor, and the variation of the metric tensor is
given by the Lie derivative 
\begin{equation}
\delta_{\xi}g_{\mu\nu}=-\nabla_{\mu}\zeta_{\nu}-\nabla_{\nu}\zeta_{\mu}=-%
\xi_{\nu}^{\alpha}\nabla_{\mu}\varepsilon_{\alpha}-\xi_{\mu}^{\alpha}%
\nabla_{\nu}\varepsilon_{\alpha}.  \label{eq:B.2}
\end{equation}
This means that the invariance of the matter action under de Sitter
``translation'' leads to the conservation law $\nabla_{\mu}\Pi^{\mu\alpha}=0$%
, where 
\begin{equation}
\Pi^{\mu\alpha}=T^{\mu\nu}\xi_{\nu}^{\alpha}=T^{\mu\alpha}-\frac{1}{4l^{2}}%
K^{\mu\alpha}.  \label{eq:B.3}
\end{equation}
This generalizes the energy-momentum tensor, combining $T^{\mu\alpha}$ with
the proper conformal current $K^{\mu\alpha}=T^{\mu\nu}\vartheta_{\nu}^{%
\alpha }$.

To obtain Einstein's equations in a locally de Sitter spacetime, consider
the variation of the Einstein-Hilbert action: 
\begin{equation}
\delta S_{g}=\frac{c^{3}}{16\pi G}\int \sqrt{-g}G_{\mu \nu }\delta g^{\mu
\nu }d^{4}x=-\frac{c^{3}}{8\pi G}\int \sqrt{-g}\nabla _{\mu }\left( G^{\mu
\nu }\xi _{\nu }^{\alpha }\right) \varepsilon _{\alpha }d^{4}x,
\label{eq:B.4}
\end{equation}%
where we have utilized Eq.~(\ref{eq:B.2}) and $G_{\mu \nu }$ denotes the
standard Einstein tensor. Local gauge invariance of this action, together
with the arbitrariness of the parameter $\varepsilon ^{\alpha }$, gives 
\begin{equation}
\nabla _{\mu }\left( G^{\mu \nu }\xi _{\nu }^{\alpha }\right) =\nabla _{\mu
} \left[ \delta _{\nu }^{\alpha }G^{\mu \nu }-\frac{1}{4l^{2}}\vartheta
_{\nu }^{\alpha }G^{\mu \nu }\right] =0.  \label{eq:B.5}
\end{equation}%
This corresponds to the Bianchi identity. As $l\rightarrow \infty $, Eq.~(%
\ref{eq:B.5}) reduces to the second Bianchi identity of General Relativity $%
\nabla _{\mu }\left( G_{(T)}^{\mu \alpha }\right) =0$; as $l\rightarrow 0$, $%
\nabla _{\mu }\left( G_{(K)}^{\mu \alpha }\right) =0$, where $G_{(K)}^{\mu
\alpha }=G^{\mu \nu }\vartheta _{\nu }^{\alpha }/4l^{2}$.

Consistency of General Relativity with de Sitter special relativity, in
which $\Pi ^{\mu \nu }$, rather than $T^{\mu \nu }$, is the conserved
current, requires generalizing Einstein's equation as \cite{60}-\cite{71} 
\begin{equation}
\mathcal{G}^{\mu \alpha }=G^{\mu \nu }\xi _{\nu }^{\alpha }=\frac{8\pi G}{%
c^{4}}T^{\mu \nu }\xi _{\nu }^{\alpha },  \label{eq:B.6}
\end{equation}%
which can be written in the form~(\ref{p9'})-(\ref{p10'}). From Eq.~(\ref%
{p10'}), as $l\longrightarrow \infty $, one recovers the usual Einstein
equations describing the dynamics of spacetime curvature; as $%
l\longrightarrow 0$ \cite{60}-\cite{71} 
\begin{equation}
G_{(K)}^{\mu \alpha }=\frac{8\pi G}{4l^{2}c^{4}}K^{\mu \alpha }.
\label{eq:B.7}
\end{equation}%
This equation describes the de Sitter effects on the geometry
of spacetime, and indicates that the energy-momentum tensor $T^{\mu \alpha }$
is the source of the dynamic curvature of General Relativity. The proper
conformal current $K^{\mu \alpha }$ is the source of the local kinematic
curvature associated with the de Sitter background.

The curvature, Ricci tensor, and the scalar curvature of the de Sitter
kinematic sector are fixed by the standard vacuum condition $G_{(K)}^{\mu
\nu }+\Lambda g^{\mu \nu }=0$ satisfied by de Sitter space (the same sign
verified independently in Sec.~\ref{sec:4.2} against $H^{2}=\Lambda /3$).
Taking the trace of this condition in four dimensions gives $%
R_{(K)}=4\Lambda $, and back-substitution gives $R_{(K)}^{\mu \nu }=\Lambda
g^{\mu \nu }$, so that 
\begin{equation}
R_{(K)\beta \mu \nu }^{\alpha }=\frac{\Lambda }{3}\left( \delta _{\mu
}^{\alpha }g_{\beta \nu }-\delta _{\nu }^{\alpha }g_{\beta \mu }\right)
\longrightarrow R_{(K)}^{\mu \nu }=\Lambda g^{\mu \nu }\longrightarrow
R_{(K)}=4\Lambda ,  \label{eq:B.8}
\end{equation}%
which is the standard, textbook curvature of de Sitter space (an Einstein
space with $R_{\mu \nu }=\Lambda g_{\mu \nu }$, $R=4\Lambda $ for $\Lambda >0
$). Substituting Eq.~(\ref{eq:B.8}) gives 
\begin{equation}
G_{(K)}^{\mu \alpha }=R_{(K)}^{\mu \alpha }-\frac{1}{2}g^{\mu \alpha
}R_{(K)}=\Lambda g^{\mu \alpha }-2\Lambda g^{\mu \alpha }=-\Lambda g^{\mu
\alpha }.  \label{eq:B.9}
\end{equation}%
Substituting Eq.~(\ref{eq:B.9}) into the splitting relation Eq.~(\ref{p10'}%
), $G_{(T)}^{\mu \alpha }-G_{(K)}^{\mu \alpha }=\frac{8\pi G}{c^{4}}(T^{\mu
\alpha }-K^{\mu \alpha }/4l^{2})$, gives directly 
\begin{equation}
G_{(T)}^{\mu \alpha }+\Lambda (x)g^{\mu \alpha }=\frac{8\pi G}{c^{4}}\left(
T^{\mu \alpha }-\frac{1}{4l^{2}(x)}K^{\mu \alpha }\right) ,  \label{eq:B10}
\end{equation}%
in agreement with Eq.~(\ref{10d}) for $\alpha =0$, obtained independently in
Sec.~\ref{sec:4.2} from the variational principle. Note that
Eq.~(\ref{eq:B.9}) identifies the kinematic sector $G_{(K)}^{\mu \alpha
}=G^{\mu \nu }\vartheta _{\nu }^{\alpha }/4l^{2}$ of Eq.~(\ref{eq:B.5}) with
the curvature of the local de Sitter background. This is an identification
(the kinematic sector is taken to be locally de Sitter), not a consequence
of Eq.~(\ref{eq:B.5}) alone. The two routes to the field equation (the direct variational derivation of Sec.~\ref{sec:4} and the Noether-current argument of this appendix) are therefore mutually consistent.

Since $K^{\mu\alpha}$ is constructed as an algebraic contraction of $%
T^{\mu\alpha}$ with the coordinate deformation tensor, the vanishing of the
material source implies the vanishing of the conformal current: $%
T^{\mu\alpha }=0$ implies $\Pi^{\mu\alpha}=0$. In the metric framework of
Refs.~\cite{60}-\cite{71}, where $l$ has no equation of its own, this
suggests that the local de Sitter kinematics is tied to the presence of
matter. In the present framework, however, $l(x)$ obeys its own field
equation, Eq.~(\ref{eq:leom}), and the value of $\Lambda(x)=3/l^{2}(x)$ is
not fixed by $K^{\mu\alpha}$: as shown in Sec.~\ref{sec:6.2}, $\Lambda(x)$
remains non-zero in vacuum and evolves dynamically, relaxing to zero only
asymptotically.

The invariance of the action under de Sitter diffeomorphisms requires that 
\begin{equation}
\nabla _{\mu }\left( G^{\mu \nu }\xi _{\nu }^{\alpha }\right) =\nabla _{\mu
} \left[ G^{\mu \nu }\left( \delta _{\nu }^{\alpha }-\frac{1}{4l^{2}(x)}%
\vartheta _{\nu }^{\alpha }\right) \right] =0,  \label{eq:B.11}
\end{equation}%
so that, 
\begin{equation}
\nabla _{\mu }G_{(T)}^{\mu \alpha }-\nabla _{\mu }G_{(K)}^{\mu \alpha
}=G^{\mu \nu }\vartheta _{\nu }^{\alpha }\partial _{\mu }\left( \frac{1}{%
4l^{2}(x)}\right) ,  \label{eq:B.12}
\end{equation}%
which shows that the right-hand side of the conservation equation is not
zero if $l(x)$ varies. This leftover term, containing the gradient of the
local radius, $\partial _{\mu }(1/4l^{2})$, is the coordinate-tensor
counterpart of the geometric source analyzed in the language of differential
forms in Sec.~\ref{sec:6}. It confirms that the spacetime
variation of $\Lambda (x)$ affects the conservation of pure dynamic curvature. To preserve gauge consistency, macroscopic spacetime must therefore couple to the conformal flow of matter and, in the extension of this work, to the axial torsion invariant.

\section{The Cartan Connection and the Gauge Potential Ambiguity}

\label{app:cartan}

The transformation~(\ref{p1}) induces a transformation in a scalar field $%
\psi (x)$ shown in Eq.~(\ref{p4}). The construction parallels the original
Poincar\'{e} gauge treatment of Kibble and Hehl et al.\ Refs.~\cite{2,3,5}.
These results hold when the transformation is global. In the case of local
symmetries, the parameters $\varepsilon ^{\mu }$ and $\lambda ^{\mu \nu }$
are spacetime-dependent, so the transformation law takes the form 
\begin{equation}
\psi ^{\prime }(x)=\left[ 1-\zeta ^{\nu }(x)\partial _{\nu }\right] \psi (x),
\label{eq:C.1}
\end{equation}%
That is, translations and rotations combine into a single general coordinate
transformation $x^{\mu }\longrightarrow x^{\prime \mu }=x^{\mu }+\zeta ^{\mu
}(x)$. From Eq.~(\ref{eq:C.1}), it follows that $\partial _{\mu }\psi $
transforms as a covariant vector field under this coordinate transformation,
generalizing Eq.~(\ref{p1}). This leads to replacing the Minkowski metric $%
\eta _{\mu \nu }$ with the spacetime-dependent metric $g_{\mu \nu }$, which
is a tensor under a general coordinate transformation \cite{79}.

Extending the scalar-field transformation law to fields that
carry a nontrivial representation of the symmetry group, such as spinors,
requires introducing the vierbein field $e^{a}{}_{\mu }(x)$, its inverse $e^{\mu }{}_{a}(x)$ and a spacetime-dependent matrix representation $(\hat{\pi}_{\mu },\hat{L}_{\mu \nu })$ of the algebra generators. The vierbein satisfies $g_{\mu \nu }=e^{a}{}_{\mu }e_{a\nu }$ and $\eta _{ab}=e_{a}{}^{\mu }e^{\nu }{}_{b}$. This is a standard construction,
paralleling the original Poincar\'{e} gauge treatment of Kibble and Hehl et
al.\ \cite{2,3,4,5}, adapted here to the local de Sitter symmetry.

We now examine the relationship between the translation gauge potential $%
h^{a}{}_{\mu }$ and the physical vierbein $e^{a}{}_{\mu }$. Writing the
unified $\mathfrak{so}(4,1)$ gauge connection 1-form as 
\begin{equation}
A_{\mu }=\frac{1}{2}\omega ^{ab}{}_{\mu }L_{ab}+\frac{1}{l(x)}h^{a}{}_{\mu
}\pi _{a},  \label{eq:C.8}
\end{equation}%
the potential $h^{a}$ transforms under a local de Sitter translation $%
\varepsilon ^{a}$ as $\delta h^{a}=-D\varepsilon ^{a}$. If the metric were
built directly as $g_{\mu \nu }=h^{a}{}_{\mu }h^{b}{}_{\nu }\eta _{ab}$, it
would therefore not be invariant under the full gauge group. The standard
resolution is the nonlinear realization of Stelle and West \cite{93}, which
was used in Ref.~\cite{94} to construct Lanczos-Lovelock actions
invariant under the (A)dS group in any dimension.

One introduces a field $y^{A}(x)$ in the vector representation of $SO(4,1)$,
with covariant derivative $Dy^{A}=dy^{A}+W^{A}{}_{B}\,y^{B}$, where $W^{AB}$
is the connection of Appendix~\ref{app:ads}. In Ref.~\cite{93} its norm is
fixed, $y^{A}y_{A}=l^{2}$, and its direction is the Goldstone field $\xi
^{a} $ of the coset $SO(4,1)/SO(3,1)$. Here we relax this constraint and
identify 
\begin{equation}
l(x)=|y|\equiv \sqrt{\eta _{AB}y^{A}y^{B}},  \label{eq:C.9}
\end{equation}%
so that the pseudo-radius is the radial mode of $y^{A}$ and $\xi ^{a}$ its
angular modes. A de Sitter gauge theory in which the de Sitter length is a
dynamical field playing the role of a dilaton was studied in Ref.~\cite{73}.
The 1-form 
\begin{equation}
E^{A}=Dy^{A}-\frac{y^{A}}{y^{2}}\,y_{B}Dy^{B},\qquad y_{A}E^{A}=0,
\label{eq:C.10}
\end{equation}%
is a spacetime 1-form and a vector under local $SO(4,1)$ transformations.
Hence $g_{\mu \nu }=\eta _{AB}E^{A}{}_{\mu }E^{B}{}_{\nu }$ is a genuine
spacetime tensor, invariant under the full gauge group. The gauge freedom
can always be used to reach the \emph{physical gauge} $y^{A}=l(x)\,\delta
_{4}^{A}$ (i.e.\ $\xi ^{a}=0$), in which 
\begin{equation}
Dy^{a}=h^{a},\qquad Dy^{4}=dl,\qquad E^{a}=h^{a},\qquad E^{4}=0.
\label{eq:C.11}
\end{equation}%
The identification of the translational gauge potential with the physical
vierbein, $e^{a}=h^{a}$, is therefore an exact gauge-fixed identity, valid
at all scales, and not a perturbative approximation. Away from the physical
gauge, the components of $E^{A}$ in the frame where $y\propto \delta
_{4}^{A} $ reproduce the nonlinear vierbein $V^{a}=O^{a}{}_{b}e^{b}+D_{z}\xi
^{a}$ of Refs.~\cite{93,94}, which transforms as a Lorentz vector with a
field-dependent parameter.

The action of the present paper can then be written in $SO(4,1)$-invariant
form. Defining the invariant 4-forms 
\begin{align}
I_{1}& =\varepsilon _{ABCDE}\,y^{E}F^{AB}\wedge Dy^{C}\wedge Dy^{D},\qquad
I_{2}=\varepsilon _{ABCDE}\,y^{E}Dy^{A}\wedge Dy^{B}\wedge Dy^{C}\wedge
Dy^{D},  \notag \\
X& =\left( y_{A}Dy^{A}\right) \wedge \left( y_{C}F^{CB}\right) \wedge Dy_{B},
\label{eq:C.12}
\end{align}%
with $\varepsilon _{01234}=+1$, one finds in the physical gauge $%
I_{1}=l\,\varepsilon _{abcd}R^{ab}e^{c}e^{d}-l^{-1}\varepsilon
_{abcd}e^{a}e^{b}e^{c}e^{d}$, $I_{2}=l\,\varepsilon
_{abcd}e^{a}e^{b}e^{c}e^{d}$ and $X=-l\,dl\wedge T^{a}\wedge e_{a}$, so that 
\begin{equation}
S_{MEHC}+S_{NY}=\int \left[ \frac{\kappa }{2|y|}I_{1}+\frac{\kappa }{4|y|^{3}%
}I_{2}+\frac{2\alpha \kappa }{|y|}X\right] .  \label{eq:C.13}
\end{equation}%
The gravitational action of Eqs.~(\ref{6}) and (\ref{eq:SNY}) is thus the
physical-gauge form of an action that is exactly invariant under
local $SO(4,1)$ transformations off shell; in particular, the Nieh-Yan
coupling does not spoil the gauge invariance. A term of the form $X$ was
considered by Westman and Z\l o\'{s}nik \cite{95,75}, where it was shown to
generate a kinetic term for $|y|$ through the torsion, in agreement with the
mechanism of Appendix~\ref{app:Seff}. The MacDowell-Mansouri invariant $%
\varepsilon _{ABCDE}\,y^{E}F^{AB}\wedge F^{CD}=l\,\varepsilon
_{abcd}F^{ab}\wedge F^{cd}$ is an independent invariant; its coefficient is
set to zero in the present work.

Finally, the limit $|y|\rightarrow 0$ is the symmetric point at which the
breaking $SO(4,1)\rightarrow SO(3,1)$ disappears. The coefficients in Eq.~(%
\ref{eq:C.13}) are singular there, so this regime, related to the conic
limit $l\rightarrow 0$ of Aldrovandi and Pereira, lies outside the domain of
the present action and is left for future work.

\section{Effective action for $l(x)$: explicit elimination of the torsion}

\label{app:Seff}

In this appendix we derive Eq.~(\ref{eq:Seff}) by eliminating the torsion 
explicitly from the action $S_{MEHC}+S_{NY}$, Eqs.~(\ref{6})
and~(\ref{eq:SNY}), without assuming a cosmological background or the form
of the solution. We consider matter that does not couple to the Lorentz
connection, ${}^{\ast }S_{ab}=0$, and use the convention of Sec.~\ref%
{sec:4.1}, $e^{a}\wedge e^{b}\wedge e^{c}\wedge e^{d}=-\varepsilon
^{abcd}\,e\,d^{4}x$ with $\varepsilon _{0123}=+1$ and indices raised with $%
\eta _{ab}$, so that $\varepsilon ^{0123}=-1$, $\varepsilon
_{abcd}\varepsilon ^{pqcd}=-2(\delta _{a}^{p}\delta _{b}^{q}-\delta
_{b}^{p}\delta _{a}^{q})$ and $\varepsilon _{abcd}\varepsilon ^{abcd}=-4!$.
Indices on $K$ and $\partial l$ are raised and lowered with $\eta _{ab}$.

\textbf{Step 1: contorsion.} Write $\omega ^{ab}=\mathring{\omega}%
^{ab}+K^{ab}$, where $\mathring{\omega}^{ab}$ is the torsion-free connection
of $e^{a}$, $de^{a}+\mathring{\omega}^{a}{}_{b}\wedge e^{b}=0$, and $%
K^{ab}=-K^{ba}=K^{ab}{}_{c}\,e^{c}$ is the contorsion 1-form, with $24$
independent components $K_{abc}=-K_{bac}$. Then 
\begin{equation}
T^{a}=K^{a}{}_{b}\wedge e^{b},\qquad R^{ab}=\mathring{R}^{ab}+\mathring{D}%
K^{ab}+K^{a}{}_{c}\wedge K^{cb}.  \label{eq:D1}
\end{equation}

\textbf{Step 2: the action in terms of $K$.} Substituting Eq.~(\ref{eq:D1})
into the curvature term of Eq.~(\ref{6}), the term linear in $K$ is a total
derivative, 
\begin{equation}
\varepsilon _{abcd}\,\mathring{D}K^{ab}\wedge e^{c}\wedge e^{d}=d\big(%
\varepsilon _{abcd}\,K^{ab}\wedge e^{c}\wedge e^{d}\big),
\end{equation}%
because $\mathring{D}e^{a}=0$ and the 3-form in brackets is a Lorentz
scalar. Dropping it, 
\begin{equation}
S_{MEHC}+S_{NY}=\int \Big(\mathring{\mathcal{L}}+\mathcal{L}_{KK}+\mathcal{L}%
_{NY}\Big)\,e\,d^{4}x,  \label{eq:D2}
\end{equation}%
where the purely Riemannian part, using $\frac{1}{2}\varepsilon _{abcd}%
\mathring{R}^{ab}\wedge e^{c}\wedge e^{d}=\mathring{R}\,e\,d^{4}x$ and $%
\varepsilon _{abcd}e^{a}e^{b}e^{c}e^{d}=4!\,e\,d^{4}x$, is 
\begin{equation}
\mathring{\mathcal{L}}=\kappa \Big(\mathring{R}-\frac{6}{l^{2}}\Big),
\label{eq:D3}
\end{equation}%
and the terms containing $K$ are 
\begin{align}
\mathcal{L}_{KK}& =\kappa \Big(K^{a}{}_{ea}\,K^{eb}{}_{b}-K^{a}{}_{eb}%
\,K^{eb}{}_{a}\Big),  \label{eq:D4} \\
\mathcal{L}_{NY}& =2\alpha \kappa \,\varepsilon ^{pcbd}\,K_{dbc}\,\partial
_{p}l.  \label{eq:D5}
\end{align}%
Eq.~(\ref{eq:D4}) follows from $\frac{\kappa }{2}\varepsilon
_{abcd}K^{a}{}_{e}\wedge K^{eb}\wedge e^{c}\wedge e^{d}$ with $\varepsilon
_{abcd}\varepsilon ^{fgcd}=-2\delta _{ab}^{fg}$; Eq.~(\ref{eq:D5}) follows
from $-2\alpha \kappa \,dl\wedge K^{a}{}_{b}\wedge e^{b}\wedge e_{a}$ with $%
dl=\partial _{p}l\,e^{p}$.

\textbf{Step 3: the contorsion equation.} The action~(\ref{eq:D2}) is
quadratic in $K$, with no derivatives of $K$: the contorsion is an auxiliary
field. Its source, Eq.~(\ref{eq:D5}), involves only the totally
antisymmetric part $K_{[abc]}$. Decomposing $K_{abc}$ into its trace,
totally antisymmetric and tensor parts, which are orthogonal under the
Lorentz-invariant quadratic form~(\ref{eq:D4}), the trace and tensor parts
vanish on shell, and the solution is purely axial, 
\begin{equation}
K_{abc}=\varepsilon _{abcd}\,A^{d}.  \label{eq:D6}
\end{equation}%
For this form both traces in Eq.~(\ref{eq:D4}) vanish, and 
\begin{equation}
\mathcal{L}_{KK}=6\kappa \,A_{d}A^{d},\qquad \mathcal{L}_{NY}=-12\alpha
\kappa \,A^{d}\,\partial _{d}l.  \label{eq:D7}
\end{equation}%
Varying with respect to $A^{d}$ gives $12\kappa (A_{d}-\alpha \,\partial
_{d}l)=0$, i.e. 
\begin{equation}
K_{abc}=\alpha \,\varepsilon _{abcd}\,\partial ^{d}l,\qquad T_{abc}=-2\alpha
\,\varepsilon _{abcd}\,\partial ^{d}l.  \label{eq:D8}
\end{equation}%
In an FLRW background, $T_{123}=-2\alpha \,\dot{l}$, which is the axial
torsion $q=2\alpha \dot{l}$ of Eq.~(\ref{eq:qalgebraic}) with $\Sigma _{A}=0$%
; in coordinate components Eq.~(\ref{eq:D8}) is the last term of Eq.~(\ref%
{eq:5.5}).

\textbf{Step 4: the effective action.} Substituting Eq.~(\ref{eq:D8}) back
into Eq.~(\ref{eq:D7}), 
\begin{equation}
\mathcal{L}_{KK}+\mathcal{L}_{NY}=6\alpha ^{2}\kappa \,(\partial
l)^{2}-12\alpha ^{2}\kappa \,(\partial l)^{2}=-6\alpha ^{2}\kappa
\,(\partial l)^{2},  \label{eq:D9}
\end{equation}%
with $(\partial l)^{2}=\eta ^{ab}\partial _{a}l\,\partial _{b}l=g^{\mu \nu
}\partial _{\mu }l\,\partial _{\nu }l$. As for any auxiliary field entering
quadratically with a linear source, the on-shell value is one half of the
source term. Collecting Eqs.~(\ref{eq:D3}) and~(\ref{eq:D9}), with $\kappa
=1/16\pi G$, 
\begin{equation}
S_{\mathrm{eff}}=\frac{1}{16\pi G}\int \sqrt{-g}\,\Big[\mathring{R}-\frac{6}{%
l^{2}}-6\alpha ^{2}\,g^{\mu \nu }\partial _{\mu }l\,\partial _{\nu }l\Big]%
+S_{m},  \label{eq:D10}
\end{equation}%
which is Eq.~(\ref{eq:Seff}) with 
\begin{equation}
Z=\frac{6\alpha ^{2}}{8\pi G},\qquad V(l)=\frac{3}{8\pi G\,l^{2}}.
\label{eq:D11}
\end{equation}%
These are the values obtained from the FLRW reduction in Sec.~\ref{sec:6.2},
which is therefore recovered here as a consequence of a covariant effective
action rather than imposed. For $l=$ const, Eq.~(\ref{eq:D10}) is the
Einstein-Hilbert action with $\Lambda =3/l^{2}$. With signature $(-,+,+,+)$
the kinetic term $-\frac{1}{2}Z(\partial l)^{2}$ has the healthy sign. The
field redefinition $\phi =\sqrt{12\alpha ^{2}\kappa }\;l$ brings it to
canonical form, $-\frac{1}{2}(\partial \phi )^{2}-V$, with $V\propto \phi
^{-2}$; for $8\pi G=1$ this is $\phi =\sqrt{6}\,\alpha \,l$, Eq.~(\ref%
{eq:phiV}).

All the expressions in this appendix were verified by computer algebra
evaluating the exterior-form action with a general, 24-component contorsion:
the linear system for $K_{abc}$ has a unique solution, which is the axial
one, Eq.~(\ref{eq:D8}).

If the matter couples to the connection, as Dirac fermions do
through their axial current, the spin density ${}^{\ast }S_{ab}\neq 0$ adds a source for $A^{d}$. Equation~(\ref{eq:D10}) then acquires the familiar four-fermion contact interaction of Einstein-Cartan theory, together with a derivative coupling $%
\partial _{\mu }l\,\bar{\psi}\gamma ^{5}\gamma ^{\mu }\psi $. These terms do
not change the counting of degrees of freedom.

\end{document}